%% file: colour_lorentz.tex
\documentclass[a4paper, aps, prd, amsfonts, floatfix, amssymb, amsmath, reprint, showkeys, nofootinbib, oneside, superscriptaddress, preprintnumbers]{revtex4-2}

\usepackage{graphicx}
\usepackage{dcolumn}
\usepackage{bm}
\usepackage{xcolor}
\usepackage{color}
\usepackage{dirtytalk}
\usepackage{soul}
\usepackage{slashed}
\usepackage{hyperref}

\renewcommand{\andname}{\ignorespaces}

\begin{document}

\preprint{ADP-24-12/T1251, DESY-24-120, LTH 1380}

\title{Transverse force distributions in the proton from lattice QCD}

\author{J.~A.~Crawford}
\affiliation{
CSSM, Department of Physics, The University of Adelaide, Adelaide SA, 5005, Australia
}
\author{K.~U.~Can}
\affiliation{
CSSM, Department of Physics, The University of Adelaide, Adelaide SA, 5005, Australia
}
\author{R.~Horsley}
\affiliation{School of Physics and Astronomy, University of Edinburgh, Edinburgh, EH9 3FD, UK}
\author{P.~E.~L~Rakow}
\affiliation{Theoretical Physics Division, Department of Mathematical Sciences, University of Liverpool, Liverpool L69 3BX, UK}
\author{G.~Schierholz}
\affiliation{Deutsches Elektronen-Synchrotron DESY, Notkestr. 85, 22607 Hamburg, Germany}
\author{H.~St\"uben}
\affiliation{Universit\"at Hamburg, Regionales Rechenzentrum, 20146 Hamburg, Germany}
\author{R.~D.~Young}
\affiliation{
CSSM, Department of Physics, The University of Adelaide, Adelaide SA, 5005, Australia
}
\author{J.~M.~Zanotti}
\affiliation{
CSSM, Department of Physics, The University of Adelaide, Adelaide SA, 5005, Australia
}

\collaboration{QCDSF Collaboration}

\date{\today}

\begin{abstract}
Single-spin asymmetries observed in polarised deep-inelastic scattering are important probes of hadron structure. The Sivers asymmetry has been the focus of much attention in QCD phenomenology and is yet to be understood at the quark level. In this Letter, we present a lattice QCD calculation of the spatial distribution of a colour-Lorentz force acting on the struck quark in a proton. We determine a spin-independent confining force, as well as spin-dependent force distributions with local forces on the order of 3 GeV/fm. These distributions offer a complementary picture of the Sivers asymmetry in transversely polarised deep-inelastic scattering. 
\end{abstract}
\maketitle

\textit{Introduction:} Deep-inelastic scattering and other hard processes provide access to the partonic structure of strongly interacting systems. Quantum Chromodynamics (QCD) is the theory that describes the strong interaction and how quarks and gluons (collectively, partons) bind together to form hadrons. The partonic behaviour of hadrons, such as the proton, is encoded in the non-perturbative dynamics of QCD. Specifically, deep-inelastic structure functions of the proton provide an intuitive lens with which to view deep-inelastic scattering data. However, the partonic picture of hadrons is incomplete. The parton model accounts for the leading order, or twist-two, contributions only, and does not incorporate any higher twist effects. In special cases, this partonic intuition can be used to understand higher-twist phenomena, such as twist-three operators being related to quark-gluon correlations \cite{Shuryak1982}. A phenomenologically interesting interpretation of these twist-three effects is that of a colour-Lorentz force \cite{Burkardt2013, Burkardt2019}. In this Letter, we present a computation of the distribution of this colour-Lorentz force in transverse impact-parameter space directly from lattice QCD. This involves extending calculations of the twist-three forward matrix element $d_2$ to off-forward kinematics and computing three new form factors which encode these force distributions. We show that these distributions provide a consistent and highly intuitive framework with which to view single-spin asymmetries in semi-inclusive deep-inelastic scattering (SIDIS) experiments. Furthermore, we reveal significant local, spin-dependent forces on the order of 3 GeV/fm, which is 3 times larger than the average force scale in QCD.
\newline

\textit{Background:} Deep-inelastic processes can be used to study the internal structure of hadrons. We consider inclusive deep-inelastic scattering (DIS) off a transversely polarised proton target,
\begin{equation}
    e^- + p \to e^- + X,
\end{equation}
where $X$ represents the unknown final states of the shattered proton. The hadronic contribution to the cross-section is described by the hadronic tensor, which can be expressed in terms of structure functions. Of interest to this scattering setup are the spin-dependent structure functions $g_1(x,Q^2)$ and $g_2(x,Q^2)$, where $x$ is the Bjorken scaling variable and $Q^2$ is the momentum transfer from the virtual photon. $g_2(x,Q^2)$ is of significant phenomenological interest as at leading order in $Q^{-2}$, it receives contributions from both twist-two and twist-three operators \cite{Blumlein1996}. Furthermore, when considering a transversely polarised target, the twist-two and twist-three contributions are of equal magnitude and hence the twist-three contributions can be reliably determined. 
\newline

Many transversely polarised DIS experiments observe surprisingly large single-spin asymmetries (SSAs) \cite{HERMES2007, HERMES2008, COMPASS2006}. Given the equal magnitude of the twist-two and twist-three operator contributions, it is expected that these asymmetries result from twist-three operators. Whilst matrix elements of twist-two operators have physical interpretations within the parton model, matrix elements of twist-three operators are instead related to quark-gluon correlations \cite{Shuryak1982}. By applying the chromodynamic lensing framework \cite{Burkardt2003}, physical interpretations of these twist-three operators can be obtained. It was argued in Ref. \cite{Burkardt2013} that these matrix elements have the semi-classical interpretation as the average colour-Lorentz force in the transverse plane acting on the struck quark in DIS. The Sivers function $f_{1T}(x, \mathbf{k}_T)$ \cite{Sivers1991} encodes the asymmetry observed in the cross-section of final states from the leading quark in SIDIS. The overall sign of the Sivers function dictates the direction of the asymmetry in final states. In a transverse momentum distribution (TMD) interpretation, the Sivers function encodes the probability distribution of finding an unpolarised quark carrying a longitudinal momentum fraction $x$ and transverse momentum $\mathbf{k}_T$ in a transversely polarised target. Whilst the Sivers function allows for tomographic scanning of the nucleon in transverse-momentum space, the approach here using colour-Lorentz forces in impact-parameter space allows for a complementary view of these asymmetries. This is because impact-parameter space is not the Fourier conjugate of the quark transverse momentum \cite{Diehl2015}.
\newline

In order to obtain a position-space density interpretation of the colour-Lorentz force \cite{Burkardt2019}, the following general matrix element is considered,
\begin{equation}
\label{eq: General ME}
    W^{\rho,\mu\nu}_{s,s'}(p,p') = \langle p',s'|\overline{q}(0)\gamma^\rho ig G^{\mu\nu}(0)q(0) |p,s \rangle,
\end{equation}
where $|p,s\rangle$ denotes a nucleon state with momentum $p$ and spin $s$, $q$ is a quark operator and $G^{\mu\nu}$ is the gluon field strength tensor. This matrix element can be parameterised in terms of eight form factors, however for the transverse force distributions, we are only interested in matrix elements of light-cone components of the operator $\overline{q}(0)\gamma^+igG^{+i}(0)q(0)$, where $i=x,y$ is the transverse index. The presence of $\gamma^+$ in the operator suggests that this force is weighted by the quark density. Light-cone coordinates can be expressed in terms of the usual Cartesian coordinates, $x^{\pm} = \frac{1}{\sqrt{2}}(x^0 \pm x^3)$. The off-forward matrix elements of the light-cone operator in Minkowski space can be expressed in terms of a minimal set of five form factors \cite{Burkardt2019},
\begin{multline}
    \label{eq: FF decomposition}
    W^{+,+i}(p,p') = \overline{u}(p',s')\bigg\{ (P^+\Delta_\perp^i - P^i\Delta^+)\gamma^+\Phi_1(t)\\
    +P^+ m_N i \sigma^{+i}\Phi_2(t) + \frac{1}{m_N}i\sigma^{+\nu}\Delta_\nu[P^+ \Delta_\perp^i \Phi_3(t)\\ - P^i\Delta^+ \Phi_4(t) ] +\frac{P^+\Delta^+}{m_N}i\sigma^{i\nu}\Delta_\nu\Phi_5(t)\bigg\}u(p,s),
\end{multline}
where $m_N$ is the nucleon mass, $P^\mu = \frac{1}{2}(p+p')^\mu$ is the average nucleon momentum, $\Delta^\mu = (p' - p)^\mu$ is the momentum transfer, $\Delta_\perp = (\Delta^x,\Delta^y)$ and $t = -\Delta^2$. In order to develop a probability interpretation in position space, we require the skewness parameter $\xi = -\frac{\Delta^+}{2P^+}$ to vanish \cite{Burkardt2000}, meaning $\Delta^+ = 0$. Hence, only the three form factors $\Phi_1(t)$, $\Phi_2(t)$ and $\Phi_3(t)$ contribute to the transverse force. We note that this decomposition is consistent with that derived in Ref. \cite{Hatta2024}, albeit their decomposition expands only to first order in $\Delta$. Finally, the two-dimensional Fourier transform of this matrix element can be interpreted as density distributions of a colour-Lorentz force in transverse impact-parameter space \cite{Burkardt2019}.
\newline

\textit{Method:} We make use of three gauge ensembles generated by the QCDSF collaboration, with three inverse coupling values of $\beta = 5.50, 5.65$ and 5.95, three lattice spacings and all lattices have an unphysical pion mass of approximately $m_\pi \approx 450$ MeV. Full lattice details can be found in the Supplemental Material, sec 1A \cite{SuppMatt}. For all ensembles, we use fermions described by a stout-smeared non-perturbatively $\mathcal{O}(a)$ improved Wilson action and a tree-level Symanzik improved gluon action \cite{Cundy2009}. The lattice spacing for all ensembles was determined using a number of singlet quantities \cite{LatticeSpacing1,LatticeSpacing2,LatticeSpacing3}. All calculations are performed at the $\text{SU}(3)$-flavour symmetric point, where the masses of all three quark flavours are set equal to the average light quark mass $\bar{m}=(m_s + 2m_l)/3$. Statistics on correlation function measurements are doubled by using two randomly generated source locations per configuration. 
\newline

We calculate matrix elements of local Euclidean operators taking the form 
\begin{equation}
\label{eq: O5 interaction}
    \mathcal{O}^{[5]}_{[i\{j]4\}} = -\frac{1}{4}\overline{q}(0)\gamma_{[i}\gamma_5\overleftrightarrow{D}_{\{j]}\overleftrightarrow{D}_{4\}}q(0),
\end{equation}
where $\overleftrightarrow{D} =\frac{1}{2}(\overrightarrow{D} - \overleftarrow{D})$ is the symmetrised covariant derivative, $\{...\}$ ($[...]$) denote (anti-)symmetrisation of indices, $i,j \in \{1, 2, 3\}$, $i \neq j$. This operator can be related to the operator in Equation \eqref{eq: General ME} through the identity $\overleftrightarrow{D}_\mu = \frac{1}{2}\{\gamma_\mu, \overleftrightarrow{\slashed{D}}\}$, the QCD equation of motion $\slashed{D}q = \overline{q}\overleftarrow{\slashed{D}}=0$, and $[D_\mu,D_\nu]=gG_{\mu\nu}$\cite{Shuryak1982}. Matrix elements of this operator are determined from ratios of three-point and two-point functions in lattice QCD. Disconnected contributions to the three-point functions are not considered at this stage. We use three source-sink separations, allowing for a two-state fit to be used to better control excited state contamination. The temporal separations across different ensembles are set to be approximately equal in fm. 
\newline

Having computed the ratios of correlators on the lattice, we apply a matching procedure to equate matrix elements of the operator $\mathcal{O}^{[5]}_{[i\{j]4\}}$ to the off-forward twist-three matrix elements in Equation \eqref{eq: FF decomposition}. Details of the ratio calculation and matching procedure can be found in the Supplemental Material sec 1B \cite{SuppMatt} (see also references \cite{Capitani1998, Gockeler2003} therein).  This allows us to construct a series of linear equations for the form factors $\Phi_i(t)$, which we then solve numerically to extract the form factors at a given $t$.
\newline

Due to the reduced symmetries available on the hypercubic lattice, $\mathcal{O}^{[5]}_{[i\{j]4\}}$ mixes with lower dimensional operators which transform under the same irreducible representation $\tau_1^{(8)}$ of the hypercubic symmetry group $H(4)$ \cite{Baake1981}. We incorporate this effect when imposing momentum subtraction (MOM)-like renormalisation conditions \cite{Martinelli1994, Gockeler1998}. Ideally, we would like to match our regularisation-independent momentum subtraction (RI$^\prime$-MOM) results to the modified minimal substraction ($\overline{\text{MS}}$) scheme, however at present, there are no available continuum perturbative calculations for these twist-three operators. Instead, we renormalise our results in the RI$^\prime$-MOM scheme at an intermediate scale, and then evolve these results to a common scale of $\mu = 2$ GeV \cite{Burger2022}. Details of the renormalisation procedure are outlined in the Supplemental Material, sec 1C \cite{SuppMatt} and references \cite{Kodaira1994, Kodaira1996, Perlt2018, Gockeler1999, Gockeler2000, Gockeler2005, Constantinou2014} therein.
\newline

\begin{figure}[h!]
    \centering
    \includegraphics[width = 8.6cm]{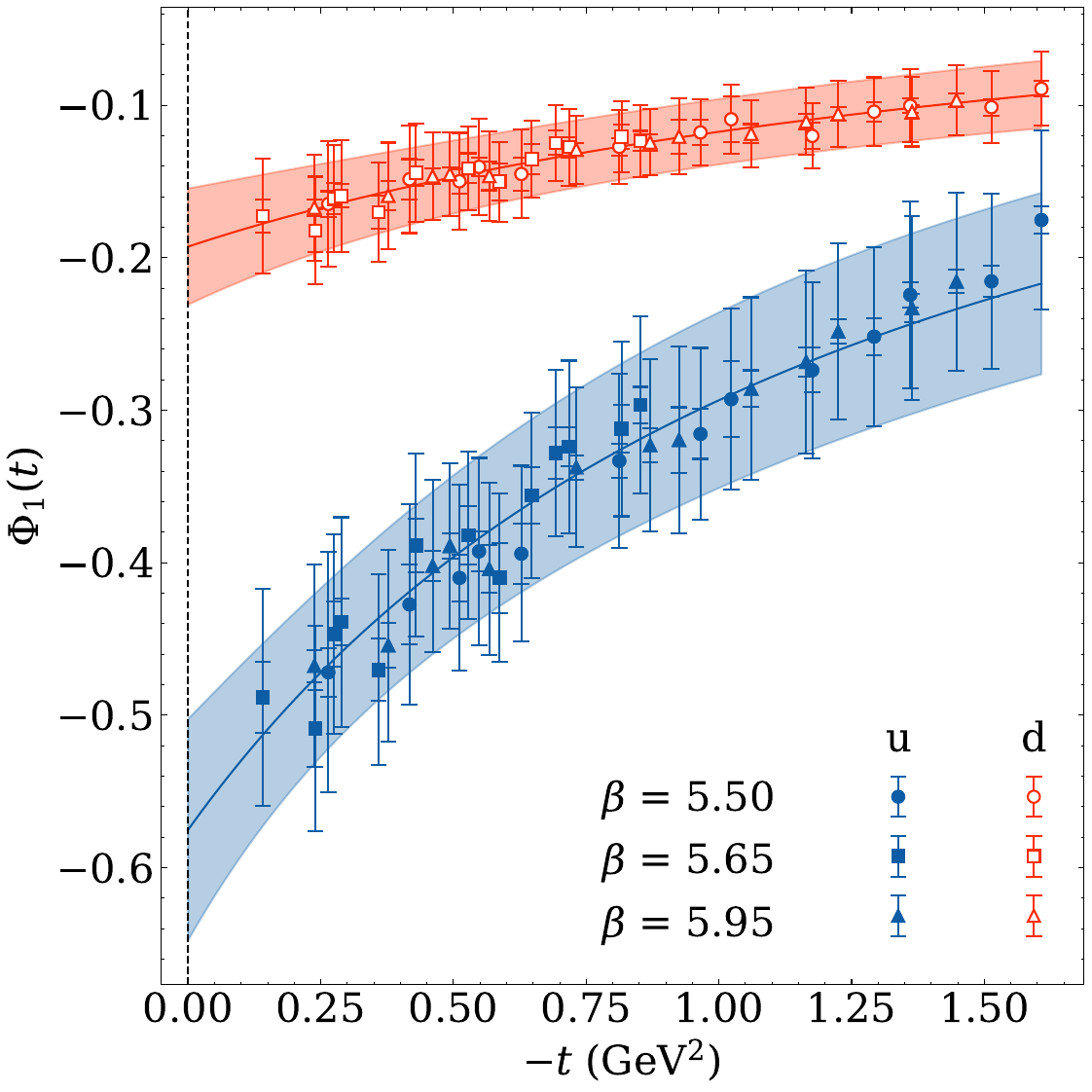}
    \caption{(Colour online) Renormalised, $\mathcal{O}(a)$ corrected results for the $\Phi_1$ form factor for the up and down quarks on all ensembles. For the fitted curves, refer to the main text.}
    \label{fig: Phi1 FF}
\end{figure}

\textit{Results:} 
In Figure \ref{fig: Phi1 FF}, we show the lattice results for the $\Phi_1$ form factor for both the up and down quarks. Discretisation artefacts have been removed from all lattice points, with the raw lattice points and translation procedure shown in the Supplemental Material, sec 1D \cite{SuppMatt}. This translation procedure increases the uncertainty in the lattice points, and so we use the inner errorbar to indicate the pure statistical uncertainty from the lattice calculation, and the outer errorbar to indicate the uncertainty after the translation. A strong, non-zero signal is seen for both quarks from all ensembles. Interestingly, the $\Phi_1$ form factor is negative for both quark flavours, for all values of the momentum transfer. This sign is indicative of a universally attractive force. Finally, we note that the magnitude of the up quark $\Phi_1$ form factor is a little more than double in size compared to the down quark, consistent with typical quark counting expectations. Allowing for small $\mathcal{O}(a)$ corrections, we fit $\Phi_1$ using a modified dipole ansatz,
\begin{equation}
\label{eq: dipole ansatz}
    \Phi_i(t) = \frac{\Phi_i(0)+b_ia}{\left(1 - t\left(\frac{1}{\Lambda_i^2}+c_ia\right) \right)^2},
\end{equation}
where $\Lambda_i$ is the dipole mass and $b_i$ and $c_i$ parameterise the $\mathcal{O}(a)$ corrections. We observe good agreement between the lattice results and the dipole fit ansatz, with a $\chi^2/d.o.f$ of 1.31 and 0.40 for the up and down quarks respectively. The $a$-dependence is not as pronounced for the smaller magnitude down quark, and so the inclusion of $a$-dependent corrections for the fit results in a degree of overfitting. 
\newline

\begin{figure}[h!]
    \centering
    \includegraphics[width = 8.6cm]{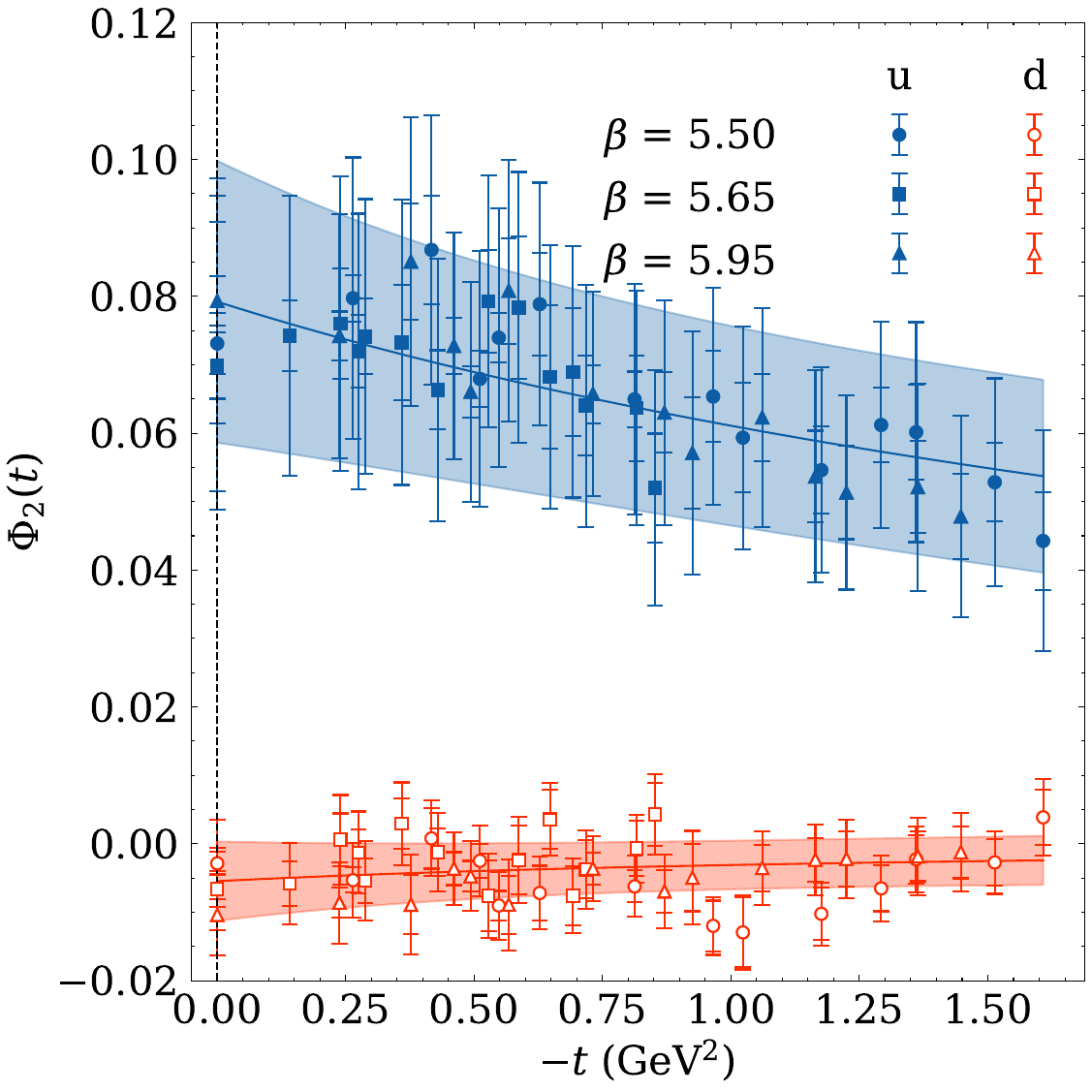}
    \caption{(Colour online) Renormalised, $\mathcal{O}(a)$ corrected results for the $\Phi_2$ form factor for the up and down quarks on all ensembles. For the fitted curves, refer to the main text.}
    \label{fig: Phi2 FF}
\end{figure}

Figure \ref{fig: Phi2 FF} shows the lattice results for the $\Phi_2$ form factor for both the up and down quarks. Whilst a non-zero signal is determined for the up quark, the down quark results are compatible with zero at this precision. Of note is the forward limit $\Phi^{(f)}_2(0) = d_2^{(f)}$. We determine a positive value for $d_2$ for the up quark and a value of $d_2$ consistent with zero for the down quark. Having removed discretisation artefacts, we compute the proton and neutron values for $d_2$, and we determine $d_2^{(p)}(m_\pi \approx 450\, \text{MeV}, a = 0) = 0.046(7)_{\rm{stat}}(16)_{\rm{sys}}$ and $d_2^{(n)}(m_\pi \approx 450\, \text{MeV}, a=0) = 0.023(5)_{\rm{stat}}(8)_{\rm{sys}}$. The artefact subtraction procedure is outlined in the Supplementary Material, sec 1D \cite{SuppMatt}. As the aim of this study is to determine the $t$-dependence of the form factors, rather than a precision measurement of $d_2$, we do not attempt a direct comparison with other lattice calculations \cite{Burger2022, Gockeler2005}, but note that our estimate is of comparible magnitude without considering quark mass effects. The results in Figure \ref{fig: Phi2 FF} show that the up quark shows some evolution in $t$ and is well approximated by a dipole function with an $a$-dependent correction to $\Phi_2(0)$ and setting $c_2 = 0$, with a $\chi^2/$d.o.f of 0.81. We note that the dipole form cannot reliably fit form factors that fluctuate about zero, as is the case for our down quark results for $\Phi_2(t)$. To bypass this issue, we instead construct and fit to the isovector ($u-d$) and isoscalar ($u+d$) combinations of the form factors. By taking a difference of the resulting fits, we are able to reconstruct a form that describes our $d$-quark form factor, as shown by the red curve in Figure \ref{fig: Phi2 FF}.
\newline

\begin{figure}[h!]
    \centering
    \includegraphics[width = 8.6cm]{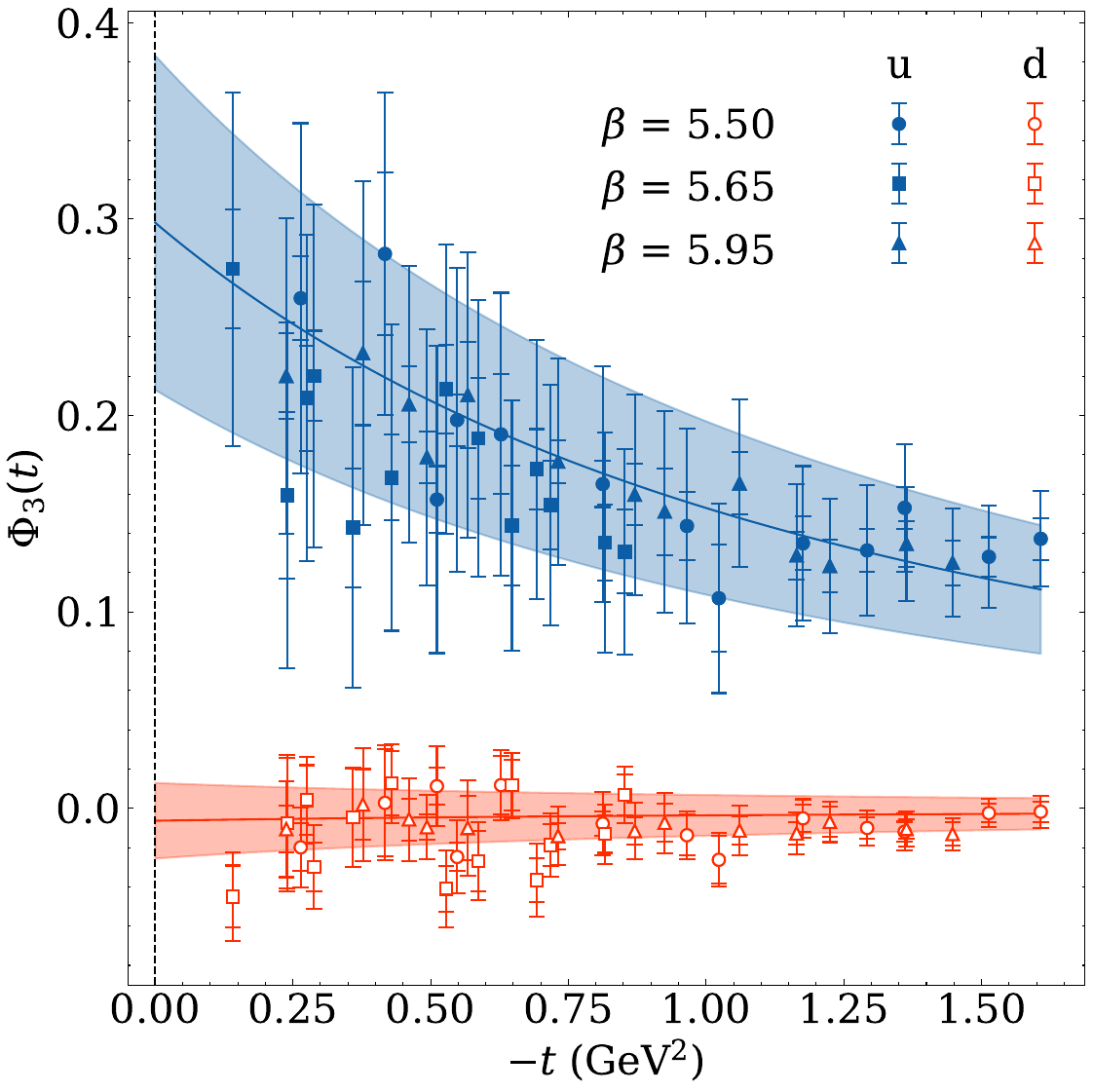}
    \caption{(Colour online) Renormalised, $\mathcal{O}(a)$ corrected results for the $\Phi_3$ form factor for the up and down quarks on all ensembles. For the fitted curves, refer to the main text.}
    \label{fig: Phi3 FF}
\end{figure}

In Figure \ref{fig: Phi3 FF}, we show the lattice results for the $\Phi_3$ form factor for both the up and down quarks. Similar to $\Phi_2$, it is more difficult to isolate a non-zero signal for the down quark, however the up quark shows a strong signal. The difference in magnitude between the two quark flavours for this form factor is quite noteworthy. Through a simple quark-diquark interpretation, the down quark is bound in a scalar diquark within the nucleon and hence its contributions to any spin-dependent observables is heavily suppressed. Using models which emphasise the diquark scenario, such as those used in Refs. \cite{Cloet2012, Wang2024}, could study these twist-three matrix elements to verify this conjecture. A dipole ansatz with an $a$-dependent correction to $\Phi_3(0)$ and $c_3 = 0$ is fit to the both sets of quark results, with the up quark results being well described by the dipole fit, having a $\chi^2/$d.o.f of 1.32. Similarly to $\Phi_2$, as the down quark results are consistent with zero at this precision, we reconstruct the down quark fit from a combination of iso-vector and iso-scalar fits.
\newline

\textit{Colour-Lorentz forces:} The two-dimensional Fourier transforms of the form factors offer a visualisation of the colour-Lorentz force in the transverse plane:
\begin{equation}
    \mathcal{F}_{s's}^j(\mathbf{b}) = \int \frac{d^2 \Delta_\perp}{(2\pi)^2} e^{-i\mathbf{b}\cdot \mathbf{\Delta}_\perp}F^j_{s's}(\mathbf{\Delta}_\perp),
\end{equation}
where
\begin{multline}
    F_{s's}^j(\mathbf{\Delta}_\perp) = \frac{i}{\sqrt{2}P^+}\\
    \times \bigg\langle p^+, \frac{\Delta_\perp}{2}, s'\bigg| \overline{q}(0)\gamma^+igG^{+j}(0)q(0) \bigg| p^+, -\frac{\Delta_\perp}{2},s \bigg\rangle.
\end{multline}
Here, $j=x,y$ is the transverse index, $s', s$ denote the nucleon polarisation and $\mathbf{\Delta}_\perp$ is the transverse momentum transferred to the nucleon, conjugate to the impact parameter $\mathbf{b}$. Utilising the form factor decomposition of the matrix element in Equation \eqref{eq: FF decomposition}, the contributions to the colour-Lorentz force can be expressed as
\begin{multline}
\label{eq: Force FT}
    \mathcal{F}_{s's}^j(\mathbf{b}) = \frac{i}{\sqrt{2}P^+}\int \frac{d^2\Delta_\perp}{(2\pi)^2}e^{-i\mathbf{b}\cdot\mathbf{\Delta}_\perp} \times\\
    \overline{u}(p',s')\bigg[ P^+ \Delta^j \gamma^+ \Phi_1(t) + P^+ m_N i\sigma^{+j}\Phi_2(t)\\
    +\frac{P^+ \Delta^j}{m_N}i\sigma^{+\nu}\Delta_\nu\Phi_3(t)\bigg]u(p,s),
\end{multline}
where, in the following, we denote the contributions to the total force from the $\Phi_i$ form factor as $\mathcal{F}_i$. 
\newline

\begin{figure}
    \centering
    \includegraphics[width = 8.6cm]{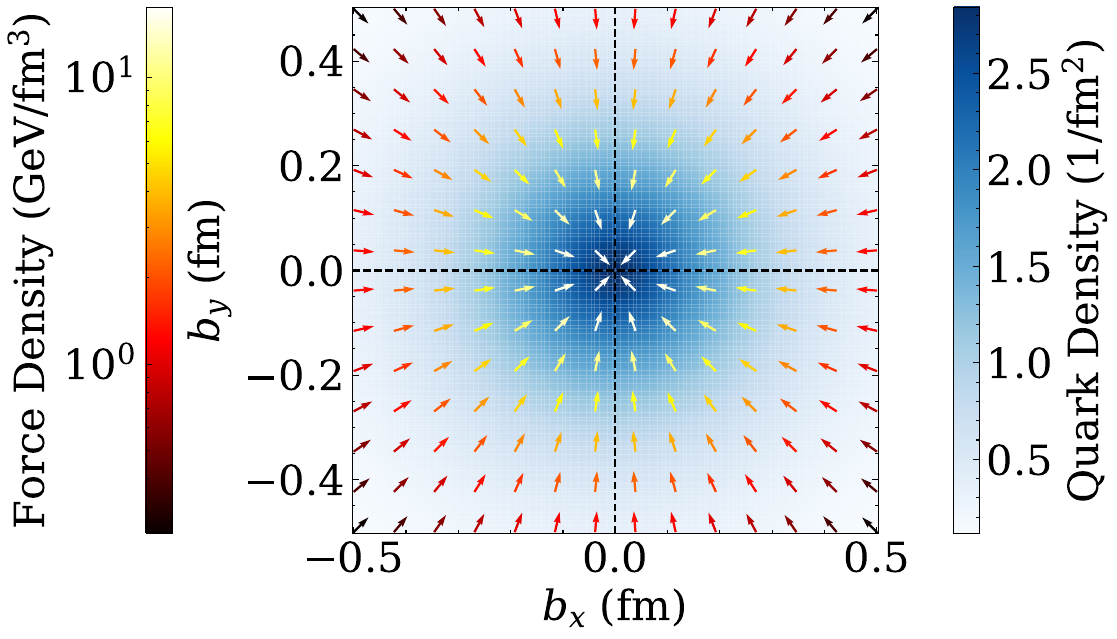}
    \caption{(Colour online) Distribution of the colour-Lorentz force acting on an unpolarised up quark in the transverse plane (indicated by the vector field) superimposed on the up quark density distribution in impact parameter space for an unpolarised proton. Vectors have unit length and their magnitude is indicated by the left colour bar.}
    \label{fig:Unpol_quark_force_density}
\end{figure}

In Figure \ref{fig:Unpol_quark_force_density}, we show the distribution of the $\mathcal{F}_1$ force on an unpolarised up quark in an unpolarised proton, using the $\mathcal{O}(a)$ improved dipole fit. This corresponds to the first term in Equation \eqref{eq: Force FT}. For the derivation of the impact parameter space representation, refer to the Supplemental Material, sec 1E \cite{SuppMatt} and reference \cite{Lorce2018} therein. The force is attractive for all regions of impact-parameter space, which is consistent with the notion of confinement. We also note that the magnitude of the weighted colour-Lorentz force density decreases towards zero as we approach the origin. As these forces are weighted by the corresponding quark densities, we divide the quark density dependence out to estimate the magnitude of these colour-Lorentz forces. We compute the quark density distributions in impact parameter space through 2D Fourier transforms of the electromagnetic form factors $F_1(t)$ and $F_2(t)$, which are computed on the same lattice ensembles as the twist-three form factors. For this, we follow the procedure outlined in Refs. \cite{QCDSF2006, Diehl2005}. We plot the obtained quark density distributions alongside the force distributions to assist in visualisation. 
\newline

By dividing the weighted force density by the quark density, we estimate local forces on the order of $\sim$ 3 GeV/fm close to the origin. This calculation is detailed in the Supplementary Material, sec 1F. It is interesting to compare this result to alternative pictures of forces in the nucleon. The static quark potential predicts a constant force of approximately 1 GeV/fm at large separations \cite{Bagan1985}. By dividing the weighted force density by the quark density for the up quark in the radial direction, we calculate a generally constant force. Recent work in Refs. \cite{Shanahan2018,Hackett2023} computes forces on the order of 0.05 GeV/fm inside the nucleon inferred from mechanical pressure distributions. These predict a repulsive pressure force close to the origin, however these distributions differ from our calculations as they are representative of a stable nucleon, whereas our result is for a struck up quark at the moment it is ejected from the proton.
\newline

\begin{figure}
    \centering
    \includegraphics[width = 8.6cm]{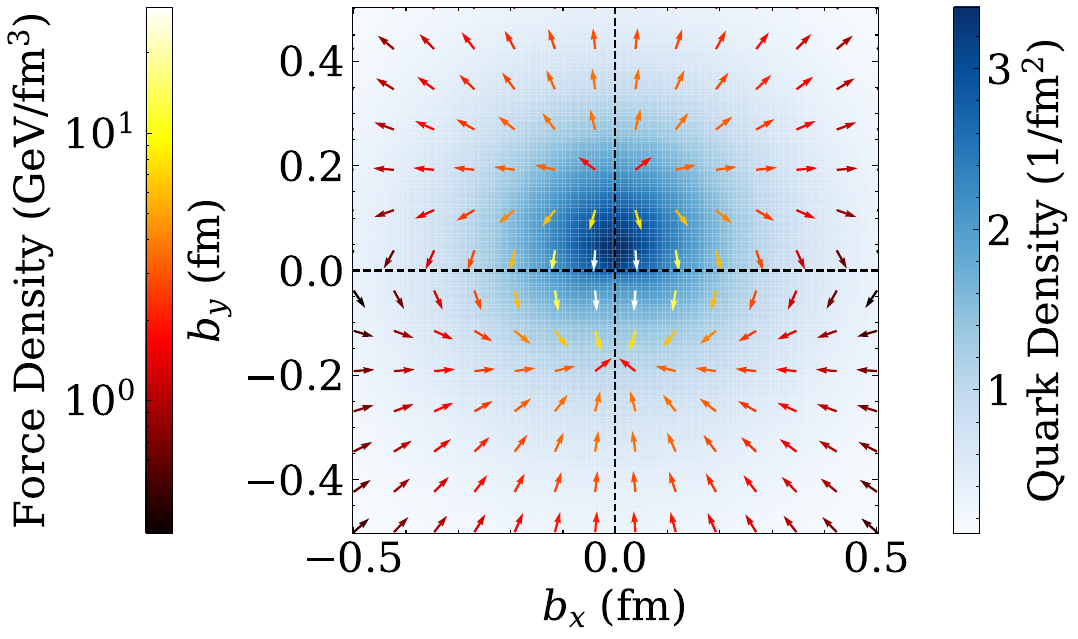}
    \caption{(Colour online) Distribution of the colour-Lorentz force acting on an unpolarised up quark superimposed on the up quark density distribution in transverse impact parameter space for a proton polarised in the $\hat{x}$ direction. Conventions are the same as Fig. \ref{fig:Unpol_quark_force_density}.}
    \label{fig:PolX_quark_force_density}
\end{figure}

In Figure \ref{fig:PolX_quark_force_density}, we show the combined distribution of the $\mathcal{F}_2$ and $\mathcal{F}_3$ forces on an unpolarised up quark in a proton polarised in the $\hat{x}$-direction, using the corresponding $\mathcal{O}(a)$ improved dipole fits. This corresponds to the second and third terms in Equation \eqref{eq: Force FT}, and the impact parameter representation is derived in the Supplemental Material, sec 1E \cite{SuppMatt}. These forces are combined as they correspond to the spin-flip matrix elements, as opposed to $\mathcal{F}_1$, which is diagonal in the nucleon spins. This force distribution is much more complex than the unpolarised case. We note that the magnitude of the vectors close to the origin is very sensitive to the fit model chosen for the form factors. We use a dipole fit for the form factors, however this results in a singularity at the origin. Other fit choices remain finite at the origin, but are still large. Beyond a distance of 0.25 fm, the magnitudes of the forces becomes model independent and are on the order of 3 GeV/fm.  The up quark is more likely to be struck in the upper half of the plane, where the force magnitude is largest and points downwards. This matches our intuition for the struck quark to be attracted back to the remnants of the shattered proton. In the Sivers asymmetry, quarks in a $+\hat{x}$-polarised proton are deflected in the $-\hat{y}$ direction, generating an asymmetric distribution of final states. Figure \ref{fig:PolX_quark_force_density} shows a significant force acting on the struck quark, pulling it in the $-\hat{y}$ direction. Therefore, the distributions produced by the Fourier transforms of these form factors are consistent with experimental observations.
\newline

\textit{Summary and Conclusions: } 
In this study, we have computed three new form factors of off-forward twist-3 matrix elements in lattice QCD and determined the resulting distribution of colour-Lorentz forces in 2D impact parameter space. We find strong signals for $\Phi_1(t)$ for both quark flavours and a strong signal for $\Phi_2(t)$ and $\Phi_3(t)$ for the up quark. Increased statistics would assist in discriminating the down quark signal for the $\Phi_2(t)$ and $\Phi_3(t)$ form factors from noise, allowing for an assessment of the flavour dependence of these forces. There is modest lattice spacing dependence in the form factors, which is more pronounced for the smaller valued $\Phi_2(t)$ form factor. Further work is required to extrapolate these results to physical pion masses to allow for direct comparison with experiment and other lattice determinations. Furthermore, expanding the momentum range to extend the range of $t$ values on which the form factors are computed would provide increased resolution of the force distributions close to the origin. This would further aid any model dependence studies of the force distributions. 
\newline

The 2D Fourier Transforms of these form factors reveal large, local forces which act on the struck quark during DIS. The force distribution for an unpolarised proton reveals a central restoring force, whilst the force distribution for a transversely polarised proton indicates a large downwards force in the region of highest quark density. To connect with phenomenology, we have shown how these results provide a complementary perspective on the Sivers asymmetry observed in transversely polarised SIDIS experiments. These 2D images of force distributions create simple and intuitive pictures of the complex phenomena of single-spin asymmetries. This work paves the way for further studies of the transverse distributions of these forces and their relationship to single-spin asymmetries observed in transversely polarised DIS experiments.

\begin{acknowledgments}
The authors would like to thank Matthias Burkardt for many useful discussions. The numerical configuration generation (using the BQCD lattice QCD program \cite{Haar2017}) and data analysis using the \texttt{CHROMA} software package \cite{Chroma2005}. Calculations were performed using the Cambridge Service for Data Driven Discovery (CSD3), the Gauss Centre for Supercomputing (GCS) supercomputers JUQUEEN and JUWELS (John von Neumann Institute for Computing, NIC, J\"ulich, Germany), resources provided by the North-German Supercomputer Alliance (HLRN), the National Computer Infrastructure (NCI National Facility in Canberra, Australia supported by the Australian Commonwealth Government), the Pawsey Supercomputing Centre, which is supported by the Australian Government and the Government of Western Australia and the Phoenix HPC service (University of Adelaide). JAC is supported by an Australian Government Research Training Program (RTP) Scholarship. RH is supported by STFC through grants ST/T000600/1 and ST/X000494/1. KUC, RDY and JMZ are supported by the Australian Research Council grant DP190100297 and DP220103098. For the purpose of open access, the authors have applied a Creative Commons Attribution (CC BY) licence to any author accepted manuscript version arising from this submission.
\end{acknowledgments}

\bibliography{colour_lorentz.bib}
\input{supp_material}

\end{document}

%% file: supp_material.tex
\renewcommand{\andname}{\ignorespaces}

\onecolumngrid
\setcounter{page}{1}

\section*{Supplemental Materials}

\subsection{Full Lattice Details}
\begingroup
\setlength{\tabcolsep}{10pt}
\renewcommand{\arraystretch}{1.5}
\begin{table*}[h!]
    \centering
    \caption{Details of gauge ensembles used in this work. $N_f$ is the number of fermion flavours, $\beta$ is the inverse gauge coupling, $c_{SW}$ is the clover coefficient, $\kappa_l$ ($\kappa_s$) is the hopping parameter for the light (strange) quarks, $L$ ($T$) is the spatial (temporal) extent of the lattice, $a$ is the lattice spacing, $m_\pi$ ($m_K$) is the pion (kaon) mass, $t_{sep}$ is the temporal separation between the source and sink and $N_{meas}$ is the number of measurements made.}
    \label{tab: Lattice details}
    \begin{tabular}{c c c c c c c c c}
    \hline\hline
        $N_f$ & $\beta$ & $c_{SW}$ & $\kappa_l$, $\kappa_s$ & $L^3\times T$ & $a$ & $m_\pi$, $m_K$ & $t_{sep}/a$ & $N_{\text{meas}}$ \\
         & & & & & (fm) & (MeV) \\
         \hline
        $2+1$\vspace{1mm} & 5.50 & 2.65 & 0.120900  & $32^3 \times 64$ & 0.074 & 465  & 11, 13, 15 & 3528\\
        
        $2+1$ & 5.65 & 2.48 & 0.122005  & $48^3\times96$ & 0.068 & 412  & 11, 14, 17 & 1074\\

        $2+1$ & 5.95 & 2.22 & 0.123460  & $48^3\times 96$ & 0.052 & 418  & 14, 18, 22 & 1014 \\
         \hline\hline
    \end{tabular}
    
\end{table*}
\endgroup

\subsection{Computation of Matrix Elements}
Matrix elements are determined from lattice two- and three-point functions of the form
\begin{equation}
\label{eq: 2pt func}
\begin{aligned}
       C_{\text{2pt}}(\mathbf{p},t) &= P_+^{\alpha\beta} C_{\text{2pt},\alpha\beta}(\mathbf{p},t), \\
       &= \sum_{\mathbf{x}}e^{-i\mathbf{p}\cdot\mathbf{x}}P_+^{\alpha\beta}\langle \chi^\beta(\mathbf{x},t) \overline{\chi}^\alpha(\mathbf{0},0) \rangle,
\end{aligned}
\end{equation}
\begin{equation}
    \label{eq: 3pt func}
    C_{\text{3pt}}(\Gamma;\mathbf{p}\,',t;\mathbf{q},\tau;\mathcal{O}) =  \sum_{\mathbf{x}_1,\mathbf{x}_2}e^{-i\mathbf{p}\,'\cdot\mathbf{x}_2}e^{i\mathbf{q}\cdot\mathbf{x}_1}
    \Gamma^{\alpha\beta}\langle \chi^\beta(\mathbf{x}_2,t)\mathcal{O}(\mathbf{x}_1,\tau)\overline{\chi}^\alpha(\mathbf{0},0)\rangle.
\end{equation}
The transfer 3-momentum is denoted by $\mathbf{q} = \mathbf{p}\,' - \mathbf{p}$. The source (sink) 3-momentum is denoted by $\mathbf{p}$ $(\mathbf{p}\,')$. The nucleon is created at the source time slice $t=0$ by the interpolating current $\overline{\chi}$ and annihilated at the sink time slice $t$. For the three-point function, a local operator current $\mathcal{O}$ is inserted at time slice $\tau$ with $t > \tau > 0$. We make use of the positive-parity projector $P_+ = (1+\gamma_4)/2$ to only consider forward-propagating states in the two-point function. The spin projection matrix $\Gamma$ is defined by $\Gamma_j = -iP_+ \gamma_j \gamma_5$ with $j=1,2,3$. Unpolarised projections were also considered by using $\Gamma_{\text{unpol}} = P_+$.\newline

The correlation functions can be related to matrix elements by inserting a complete set of states. In the large Euclidean time limit, excited states are exponentially suppressed and the correlation functions can be reasonably approximated by the ground-state contribution alone,
\begin{equation}
    C_{\text{2pt}}(\mathbf{p},t) \approx \sum_{s} P_+^{\alpha\beta}\langle{0}|{\chi^\beta}|{p,s}\rangle
    \langle{p,s}|{\overline{\chi}^\alpha}|{0}\rangle\frac{e^{-E_{\mathbf{p}}t}}{2E_{\mathbf{p}}},
\end{equation}
\begin{equation}
    C_{\text{3pt}}(\Gamma;\mathbf{p}\,',t;\mathbf{q},\tau;\mathcal{O}) \approx \sum_{s,s'} \frac{e^{-E_{\mathbf{p}\,'}(t-\tau)}}{2E_{\mathbf{p}\,'}}\frac{e^{-E_{\mathbf{p}}t}}{2E_{\mathbf{p}}}\times\\
    \Gamma_j^{\alpha\beta} \langle{0}|{\chi^\beta}|{p',s'}\rangle\langle{p',s'}|{\mathcal{O}}|{p,s}\rangle\langle{p,s}|{\overline{\chi}^\alpha}|{0}\rangle.
\end{equation}
The overlap matrix elements can be written in terms of spinors,
\begin{equation}
    \langle{0}|{\chi^\alpha}|{p,s}\rangle = \sqrt{Z_{\mathbf{p}}}u^\alpha(p,s),
\end{equation}
where $Z_{\mathbf{p}}$ is the momentum-dependent overlap factor. Similarly, for the matrix element of the inserted operator $\mathcal{O}$, 
\begin{equation}
\label{eq: spinor representation}
    \langle{p',s'}|{\mathcal{O}}|{p,s}\rangle = \overline{u}(p',s')\mathcal{J}[\mathcal{O}]u(p,s),
\end{equation}
where $\mathcal{J}[\mathcal{O}]$ represents the parameterisation of the operator $\mathcal{O}$ in terms of Dirac structures. Through use of spinor identities, we can express the correlation functions as towers of exponentials,
\begin{equation}
    C_{\text{2pt}}(\mathbf{p},t) = Z_{\mathbf{p}}\frac{E_{\mathbf{p}}+m}{E_{\mathbf{p}}}e^{-E_{\mathbf{p}}t} + ...
\end{equation}
\begin{equation}
    C_{\text{3pt}}(\Gamma;\mathbf{p}\,',t;\mathbf{q},\tau;\mathcal{O}) = \sqrt{Z_{\mathbf{p}}Z_{\mathbf{p}\,'}}e^{-E_{\mathbf{p}\,'}(t-\tau)}e^{E_{\mathbf{p}}\tau}B_{00} + ...
\end{equation}
where $B_{00}$ is the ground state matrix element and is given by
\begin{equation}
\label{eq: B_00}
    B_{00} = \frac{1}{4E_{\mathbf{p}'}E_\mathbf{p}}\text{Tr} \big[\Gamma \left(-i\slashed{p}\,' + m \right)\mathcal{J}[\mathcal{O}]\left(-i\slashed{p} + m \right)\big].
\end{equation}
\newline

\begin{figure}[t]
\centering
\begin{minipage}[t]{.49\textwidth}
  \centering
  \includegraphics[width=\textwidth]{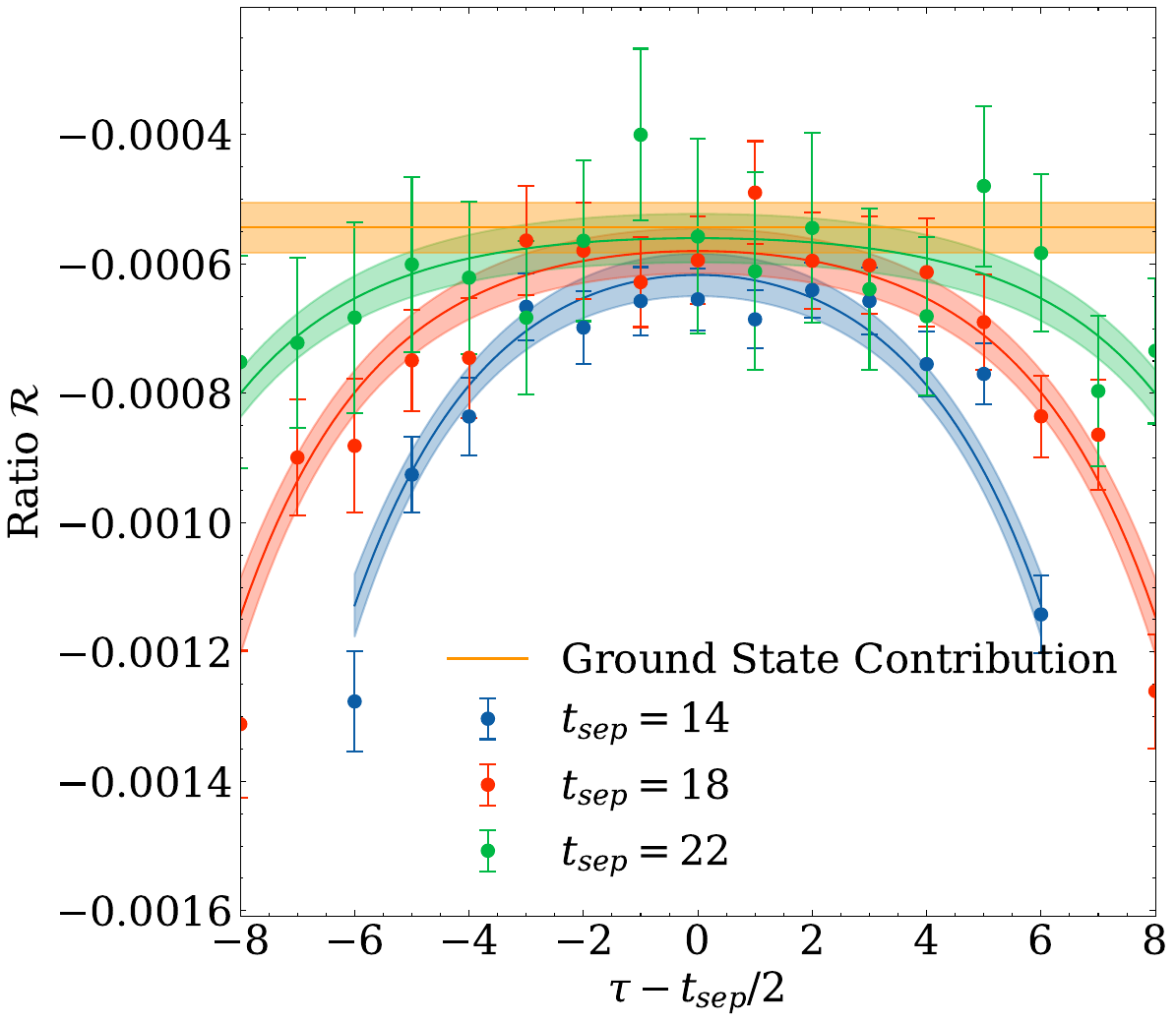}
  \caption{Example of an excited state fit in the forward limit using a bare operator from the $\beta=5.95$ ensemble. Kinematics used are $\mathbf{q} = (0,0,0)$, $\Gamma = \Gamma_2$, with operator $\mathcal{O}^{[5]}_{[2\{1]4\}}$ inserted at time slice $\tau$.}
  \label{fig: sm_t3ME}
\end{minipage}%
\hfill
\begin{minipage}[t]{.49\textwidth}
  \centering
  \includegraphics[width=\textwidth]{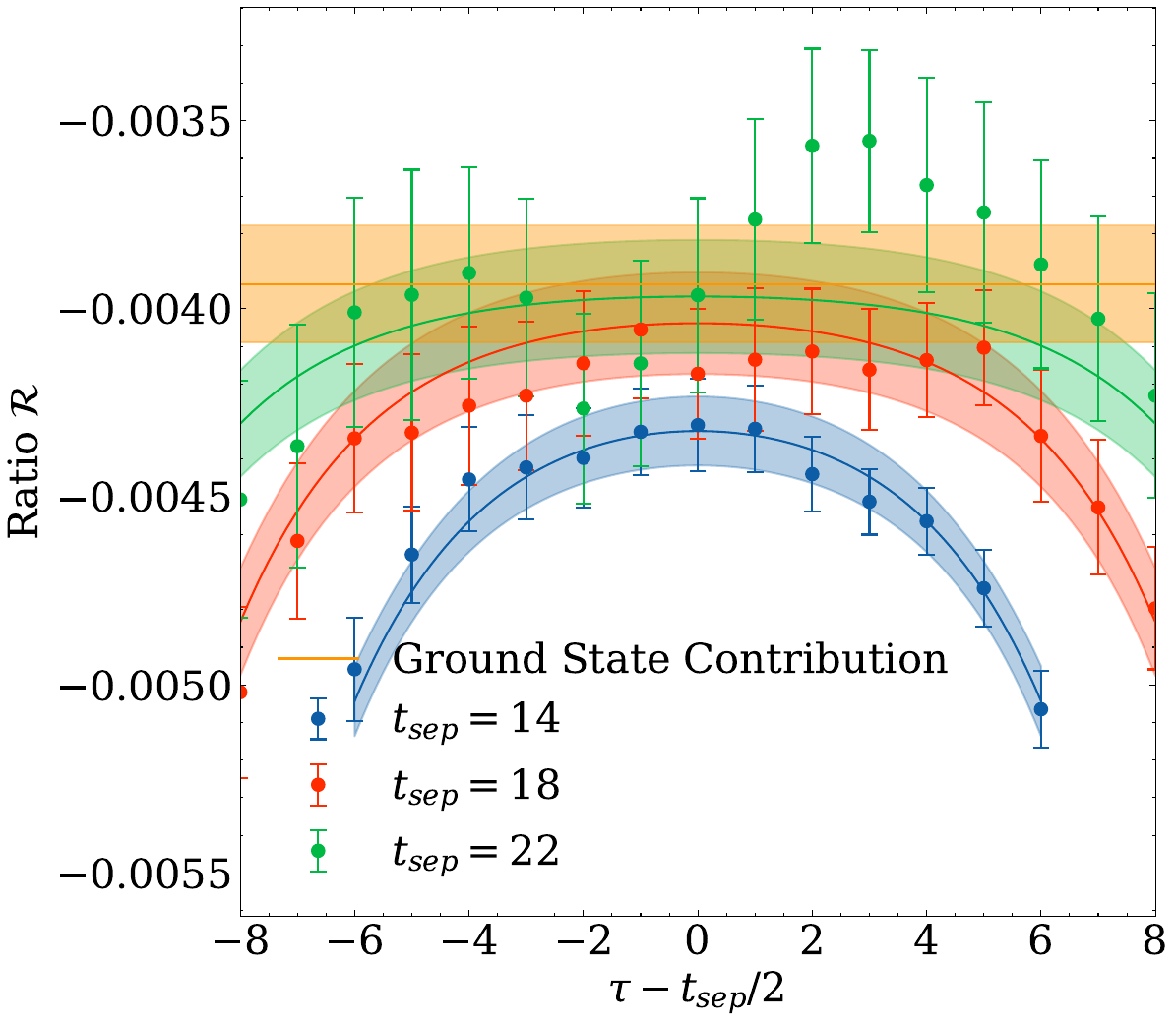}
  \caption{Comparative excited state fit of the mixing ratio $\mathcal{O}^\sigma_{[2\{1]4\}}$ at the same kinematics as Figure \ref{fig: sm_t3ME}.}
  \label{fig: sm_mixME}
\end{minipage}
\end{figure}

We compute the matrix elements of the electromagnetic and twist-three operators by computing ratios of three- and two-point correlation functions \cite{Capitani1998, Gockeler2003},
\begin{equation}
    \mathcal{R} = \frac{C_{\text{3pt}}(\Gamma_j; \mathbf{p}',t;\mathbf{q},\tau;\mathcal{O})}{C_{\text{2pt}}(\mathbf{p}',t)}  \bigg[\frac{C_{\text{2pt}}(\mathbf{p}',t)C_{\text{2pt}}(\mathbf{p}',\tau)C_{\text{2pt}}(\mathbf{p},t-\tau)}{C_{\text{2pt}}(\mathbf{p},t)C_{\text{2pt}}(\mathbf{p},\tau)C_{\text{2pt}}(\mathbf{p}',t-\tau)} \bigg]^{\frac{1}{2}},\\
\end{equation}
If one assumes ground-state dominance in the large Euclidean time limit, the ratio is proportional to the matrix element of interest,
\begin{equation}
    \mathcal{R} \stackrel{t \gg \tau \gg 0}{=} \sqrt{\frac{E_{\mathbf{p'}}E_{\mathbf{p}}}{(E_{\mathbf{p'}}+m)(E_{\mathbf{p}}+m)}}\langle p', s'|\mathcal{O}^{(q)} |p,s\rangle.
\end{equation}
However, we observe very noisy signals under this assumption, and so to control excited state contamination, we utilise a two state fit ansatz for the correlators. In order to isolate the ground state signal, we include the first excited state in our spectral decomposition ansatz of the correlation functions.
\begin{equation}
    \label{eq: ES 2pt}
    C_{\text{2pt}}(\mathbf{p},t) \approx Z_{\mathbf{p}}\frac{E_{\mathbf{p}}+m}{E_{\mathbf{p}}}e^{-E_{\mathbf{p}}t}\left(1 + A e^{-\Delta E_{\mathbf{p}}t} \right),
\end{equation}
\begin{multline}
    \label{eq: ES 3pt}
    C_{\text{3pt}}(\Gamma;\mathbf{p}\,',t;\mathbf{q},\tau;\mathcal{O}) \approx \sqrt{Z_{\mathbf{p}\,'}Z_{\mathbf{p}}}\,e^{-E_{\mathbf{p}\,'}(t-\tau)}e^{-E_{\mathbf{p}}\tau}\\ \times
    (B_{00} + B_{10}e^{-\Delta E_{\mathbf{p}\,'}(t-\tau)} + B_{01}e^{-\Delta E_{\mathbf{p}}\tau} + B_{11}e^{-\Delta E_{\mathbf{p}\,'}(t-\tau)}e^{-\Delta E_{\mathbf{p}}\tau}).
\end{multline}
The energy gap between the ground state and first excited state for a nucleon with 3-momentum $\mathbf{p}$ $\big( \mathbf{p}\,'\big)$ is denoted by $\Delta E_{\mathbf{p}}$ $\big( \Delta E_{\mathbf{p}\,'} \big)$. Figure \ref{fig: sm_t3ME} depicts an example of the fitting procedure used to extract the ground state contribution to the ratio on the $\beta = 5.95$ ensemble. The operator $\mathcal{O}^{[5]}_{[2\{1]4\}}$ is inserted into a up quark line in a proton polarised in the $\hat{y}$ direction at time slice $\tau$ with finite sink momenta $\mathbf{p}'=(1,0,0)$ and zero momentum transfer such that $\mathbf{p}=\mathbf{p}'$. The curves represent the averaged fit values, the shaded regions represent the $1\sigma$ uncertainty after fitting over 200 bootstrapped samples. From this figure, it is evident that the noise from the operator would have introduced additional uncertainty if a single-state fit were used, and that an two-state fit has improved the resolution of the matrix element. The fit parameters for the two-point functions are determined by fitting directly to the two-point correlators prior to fitting the ratio, allowing the determination of the ground state energies and energy gaps at the source and sink momenta. The remaining fit parameters lie in the three-point function and are determined by fitting directly to the ratio. For comparison, we also show the contribution of the mixing operator to the ratio in Figure \ref{fig: sm_mixME}. We note that whilst the mixing matrix element is an order of magnitude larger than the twist-3 matrix element of interest, as the mixing ratio is on the order of $10^{-2}$, see Section B, overall it represents a small correction to the value of the twist-3 matrix element.
\newline

Following from the ratio calculation, we must then match the ground state contribution to the appropriate matrix element. A complication arises as the lattice results are expressed in Euclidean space, while the matrix elements of interest are expressed in Minkowski space, and so we Wick rotate matrix elements of our lattice operator back to Minkowski space. We begin with the following representation of our $\mathcal{O}^{[5]}$ operator in Euclidean space,
\begin{equation}
\label{eq: app Euc operator}
    \mathcal{O}^{[5]}_{[\sigma\{\mu_1]\mu_2\}} = -\frac{g}{6}\overline{q}\big(\Tilde{G}_{\sigma\mu_1}\gamma_{\mu_2} + \Tilde{G}_{\sigma\mu_2}\gamma_{\mu_1} \big)q - \text{traces}.
\end{equation}
We restrict our attention to operators of the form $\mathcal{O}^{[5]}_{[i\{j]4\}}$, for $i,j=1,2,3$ and $i\neq j$. We note the Wick rotations for the dual gluon field strength tensor,
\begin{equation}
\begin{split}
    \Tilde{G}_{ij}^{(E)} &= -i\Tilde{G}_{(M)}^{ij},\\
    \Tilde{G}_{4i}^{(E)} &= \Tilde{G}_{(M)}^{0i},\\
\end{split}
\end{equation}
and for the gamma matrices,
\begin{equation}
    (\gamma^0_{(M)},\gamma^i_{(M)}) \to (\gamma_i^{(E)}, \gamma_4^{(E)}) = (-i\gamma^i_{(M)}, \gamma^0_{(M)}).
\end{equation}
Substituting the relevant Wick rotations into Equation \eqref{eq: app Euc operator}, we have
\begin{equation}
    \mathcal{O}_{[i\{j]4\}} \to -\frac{g}{6}\overline{q}\big(-i\Tilde{G}^{ij}_{(M)}\gamma_{(M)}^0 - i\Tilde{G}_{(M)}^{i0}\gamma_{(M)}^j \big)q - \text{traces}.
\end{equation}
We will now drop the $(M)$ subscript and ignore the traces from here. Using the definition of the dual field strength tensor, we have
\begin{equation}
        \mathcal{O}^{[5]}_{[i\{j]k\}} = \frac{g}{6}\overline{q}\big(i\epsilon^{ijk0}G_{k0}\gamma^0 + i\epsilon^{i0jk}G_{jk}\gamma^j \big)q.
\end{equation}
Permuting the Levi-Civita tensors and factoring them out,
\begin{equation}
    \mathcal{O}_{[i\{j]k\}}^{[5]} = -\frac{\epsilon^{0ijk}}{6}\overline{q}\big( \gamma^0igG_{k0} + \gamma^j igG_{jk}\big)q.
\end{equation}
Raising the indices of the gluon field strength tensors,
\begin{equation}
\label{eq: op matching}
    \mathcal{O}^{[5]}_{[i\{j]k\}} = \frac{\epsilon^{ijk}}{6}\overline{q}\big(\gamma^0 ig G^{k0} + \gamma^j ig G^{kj} \big)q,
\end{equation}
where we have used the anti-symmetry of the field strength tensor and $\epsilon^{0ijk}=\epsilon^{ijk}$, where $\epsilon^{ijk}$ is the three-dimensional Levi-Civita tensor. Finally, associating the terms in Equation \eqref{eq: op matching} with the matrix elements of the form
\begin{equation}
    W^{\rho,\mu\nu} = \langle{p',s'}|{\overline{q}\gamma^\rho ig G^{\mu\nu}q}|{p,s}\rangle,
\end{equation}
we have the matching
\begin{equation}
    \langle{p',s'}|{\mathcal{O}^{[5]}_{[i\{j]k\}}}|{p,s}\rangle = \frac{\epsilon^{ijk}}{6}(W^{0,k0} + W^{j,kj}).
\end{equation}

\subsection{Lattice Operators and Renormalisation}
In the case where the lattice operator is multiplicatively renormalisable, the renormalised operator $\mathcal{O}(\mu)$ is related to the bare operator $\mathcal{O}(a)$ by
\begin{equation}
    \mathcal{O}(\mu) = Z_{\mathcal{O}}(a\mu)\mathcal{O}(a),
\end{equation}
where $a$ is the lattice spacing. Calculations will be performed in Euclidean space unless otherwise stated. Due to the effects of operator mixing on the lattice, we must account for the mixing when computing matrix elements of the twist-three operator. This further obscures an already noisy signal. Operator mixing for operators relevant to the computation of moments of $g_2(x,Q^2)$ has been studied in continuum perturbation theory, notably in Refs. \cite{Kodaira1994, Kodaira1996}, which are most similar to the methods applied on the lattice.  For an operator of the form $\mathcal{O}^{[5]}_{[i\{j]4\}}$ as defined in the main text, using the massless equations of motion for QCD one can rewrite the twist-three operators in the form
\begin{equation}
    \mathcal{O}^{[5]}_{[i\{j]4\}} = -\frac{1}{4}\overline{q}\gamma_{[i}\gamma_5\overleftrightarrow{D}_{\{j]}\overleftrightarrow{D}_{4\}}q.
\end{equation}
On the lattice, the following operators, of dimension four and five respectively, would mix
\begin{equation}
    \mathcal{O}^\sigma_{[i\{j]4\}} := \frac{i}{12}\overline{q}\left(\sigma_{jk}\overleftrightarrow{D}_j - \sigma_{4k}\overleftrightarrow{D}_4 \right)q,
\end{equation}
\begin{equation}
    \mathcal{O}^0_{[i\{j]4\}} := \frac{1}{12}\overline{q}\left(\gamma_j \overleftrightarrow{D}_{[k}\overleftrightarrow{D}_{j]} - \gamma_4\overleftrightarrow{D}_{[k}\overleftrightarrow{D}_{4]}\right)q,
\end{equation}
where $i,j,k \in 1,2,3$ and $i \neq j \neq k$, $\overleftrightarrow{D}_\mu = \frac{1}{2}(\overrightarrow{D}_\mu - \overleftarrow{D}_\mu)$ and $\sigma_{\mu\nu} = \frac{i}{2}[\gamma_\mu,\gamma_\nu]$. The operator $\mathcal{O}^0_{[i\{j]4\}}$ mixes with $\mathcal{O}^{[5]}_{[i\{j]4\}}$ with a coefficient of $g^2$ and therefore vanishes in the tree-level approximation between quark states. However, the operator $\mathcal{O}^\sigma_{[i\{j]4\}}$ contributes with a coefficient proportional to $a^{-1}$ \cite{Gockeler2005} and therefore must be included in the renormalisation of our operator $\mathcal{O}^{[5]}_{[i\{j]4\}}$. Hence the renormalisation of the $\mathcal{O}^{[5]}$ operator, with indices suppressed, can be expressed as
\begin{equation}
\label{eq: Unfactored renorm}
    \mathcal{O}^{[5]}_R(\mu) = Z^{[5]}(a\mu) \mathcal{O}^{[5]}(a) + \frac{1}{a}Z^{\sigma}(a\mu) \mathcal{O}^\sigma(a),
\end{equation}
where the renormalisation constant $Z^{[5]}$ and mixing coefficient $Z^\sigma$ are to be determined by imposing the (MOM-like) renormalisation conditions \cite{Martinelli1994, Gockeler1998},
\begin{equation}
    \label{eq: RI-MOM condition 1}
    \frac{1}{12}\text{Tr} \left[ \Gamma^{[5]}_R(p) \left(\Gamma_{\text{tree}}^{[5]}(p)\right)^{-1}\right]_{p^2=\mu^2}=1,
\end{equation}
\begin{equation}
    \label{eq: RI-MOM condition 2}
    \frac{1}{12}\text{Tr} \left[ \Gamma^{[5]}_R(p) \left(\Gamma_{\text{tree}}^{\sigma}(p)\right)^{-1}\right]_{p^2=\mu^2}=0,  
\end{equation}
where the vertex function is the amputated Greens function $\Gamma(p,p,\mathcal{O}) = S^{-1}(p)G(p,p,\mathcal{O})S^{-1}(p)$, where $S^{-1}(p)$ is the inverse fermion propagator with momentum $p$, and $G(p,p,\mathcal{O})$ is the three-point correlation function with source and sink momentum $p$, and inserted operator $\mathcal{O}$.\newline

\begin{figure}[t]
\centering
\begin{minipage}[t]{.49\textwidth}
  \centering
    \includegraphics[width = \textwidth]{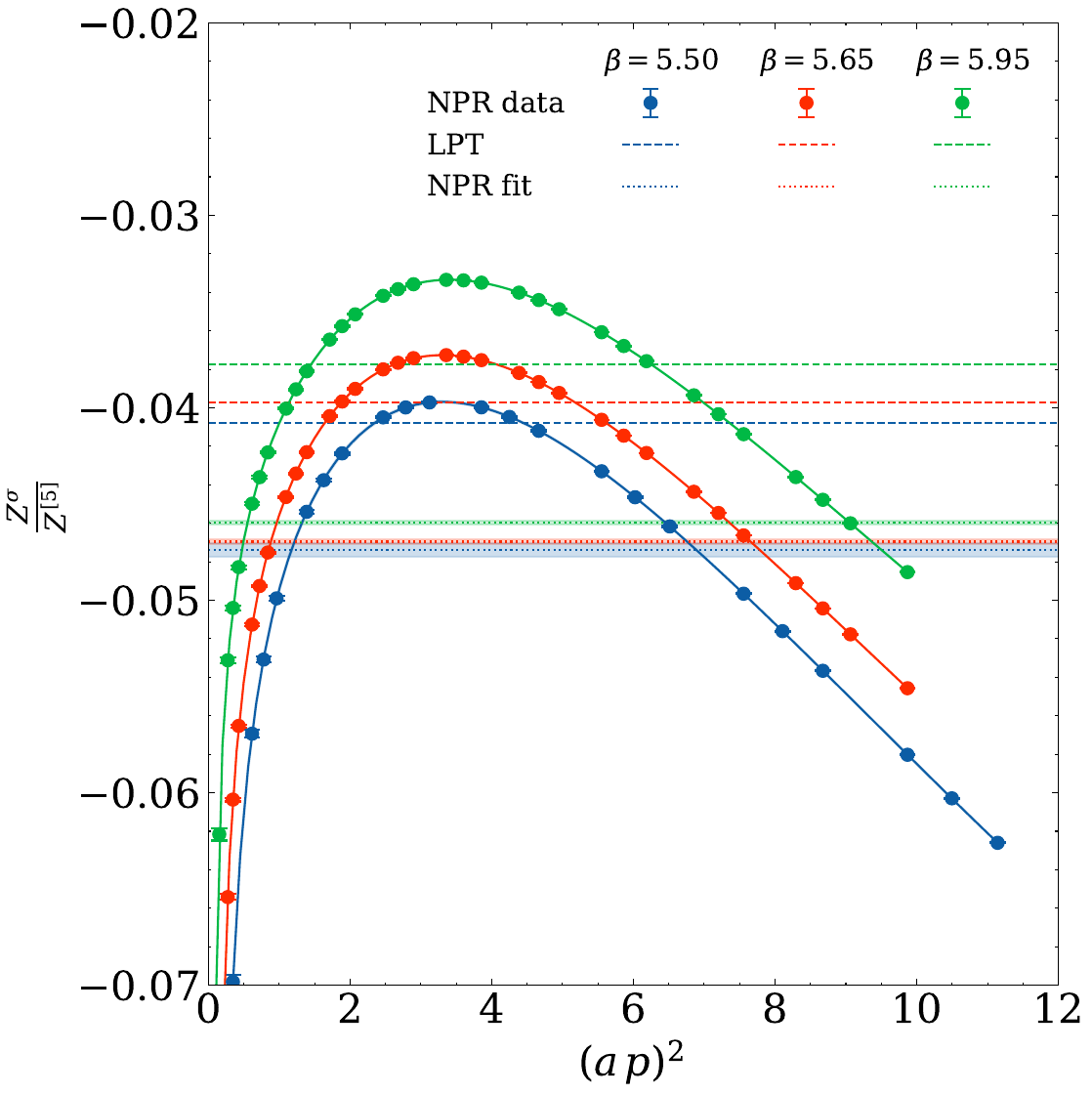}
    \caption{Comparison of the mixing coefficient $Z^\sigma/Z^{[5]}$ using both Eq. \ref{eq: LPT mixing} and a non-perturbative calculation on all three ensembles.}
    \label{fig: sm_Zsig}
\end{minipage}%
\hfill
\begin{minipage}[t]{.49\textwidth}
    \centering
    \includegraphics[width = \textwidth]{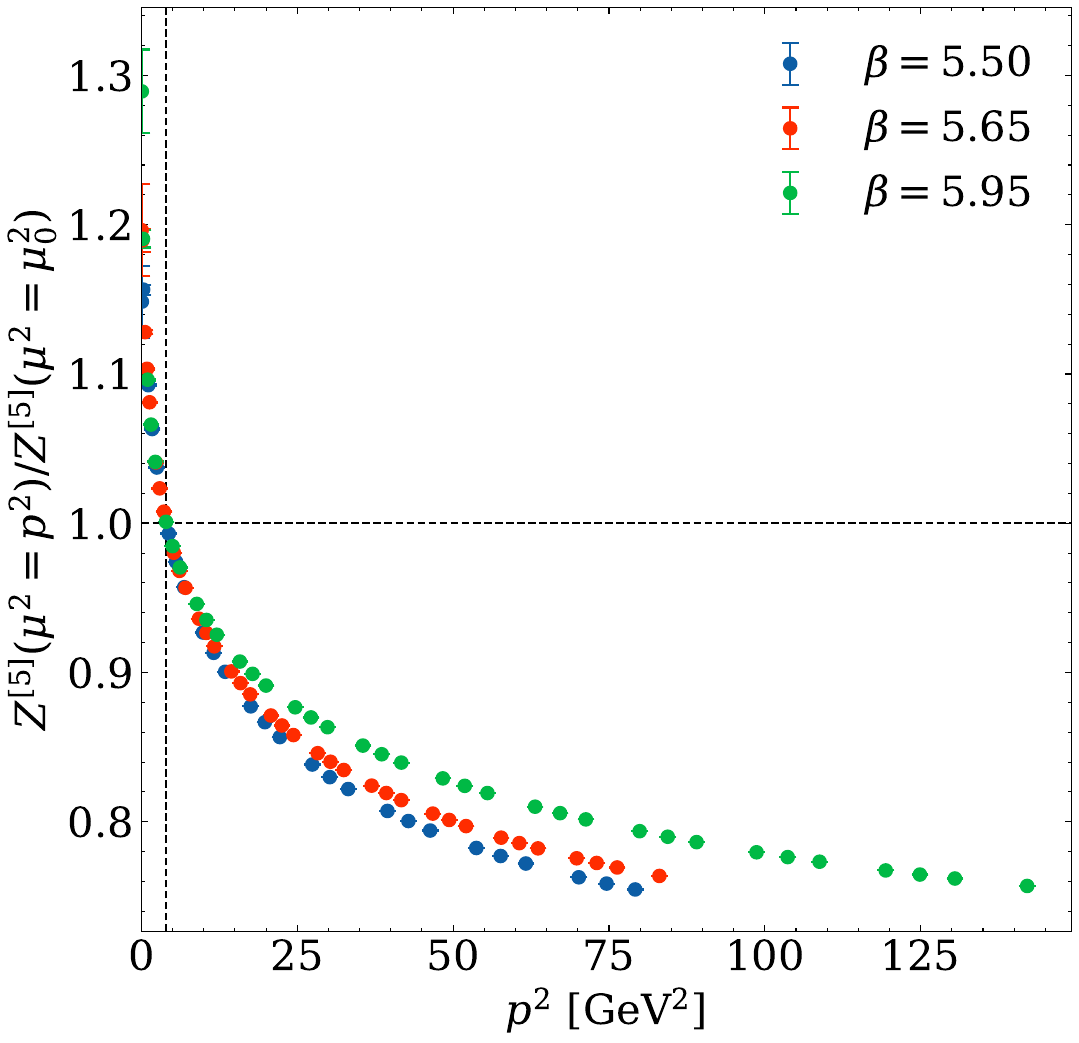}
    \caption{Calculation of the ratio $R(\mu,\mu_0)$ on all three ensembles, with $\mu_0 = 2$ GeV.}
    \label{fig: Z5 ratio}
\end{minipage}
\end{figure}

By factoring out $Z^{[5]}$ from Equation \eqref{eq: Unfactored renorm}, we observe that the operator $\mathcal{O}^{[5]}(\mu)$ will have multiplicative scale dependence if the ratio $Z^\sigma(a\mu)/Z^{[5]}(a\mu)$ is constant. We have computed the ratio $Z^{\sigma}(a\mu)/Z^{[5]}(a\mu)$ for stout link improved non-perturbative clover (SLiNC) fermions to one-loop using lattice perturbation theory (LPT), and it takes the form \cite{Perlt2018}
\begin{multline}
    \label{eq: LPT mixing}
    \frac{Z^\sigma(a\mu)}{Z^{[5]}(a\mu)}=-\frac{1}{16\pi^2}C_F g^2(11.785-44.841s_t -9.537s_t^2 
    \\+c_{SW}(-6.322+21.047s_t -20.665s_t^2)
    + c_{SW}^2(-0.206+0.797s_t -0.605s_t^2)),
\end{multline}
where $s_t=0.1$ is the stout smearing parameter used in the SLiNC action, $c_{SW}=1$ is the tree-level clover term, $C_F=4/3$ and $g^2=10/\beta$. For the three $\beta$ values used in this study, Equation \eqref{eq: LPT mixing} gives approximately $-0.0377$, $-0.0397$ and $-0.0408$ for $\beta = 5.95$, $\beta = 5.65$ and $\beta=5.50$ respectively. In Figure \ref{fig: sm_Zsig}, we show a comparison of the mixing coefficient $Z^\sigma/Z^{[5]}$ computed using the LPT result from Equation \eqref{eq: LPT mixing} and non-perturbatively using Equation \eqref{eq: RI-MOM condition 2}. We use the same lattice ensembles as in Table 1 of the main text, however in order to improve the resolution of our calculation, we make use of a twisted boundary condition for the quark fields, using the same procedure outlined in Ref. \cite{Constantinou2014}. At large $(a\,p)^2$, we note a linear trend in $Z^\sigma/Z^{[5]}$, which is typical of lattice discretisation effects entering the calculation. In the calculation of the twist-three matrix elements, we compute $Z^\sigma/Z^{[5]}$ directly from the non-perturbative data by fitting a truncated power series in $(a\,p)^2$ with a divergent $(a\,p)^{-2}$ term,
\begin{equation}
    \frac{Z^{\sigma}}{Z^{[5]}} = \frac{A_0}{(a\,p)^2} + B + \sum_{n=1}^N C_n (a\,p)^{2n},
\end{equation}
where $A_0, B$ and $C_n$ are constants to be determined from the fit. We choose $N = 5$, resulting in an optimised $\chi^2/d.o.f$ for each of the fits. As has been previously seen \cite{Gockeler1999, Gockeler2000, Gockeler2005, Burger2022}, the mixing coefficient is expected to be constant at large $(a\,p)^2$ and so we remove the lattice discretisation errors by taking the value of the mixing coefficient to be equal to the constant term $B$. Our non-perturbative estimates for the mixing coefficient are then $-0.0460(1), -0.0470(1)$ and $-0.0474(3)$ for the $\beta = 5.95$, $\beta = 5.65$ and $\beta = 5.50$ ensembles respectively. The quoted uncertainty in the NPR mixing coefficients is statistical only. We make use of these values when computing our required matrix elements, but note their deviation from the LPT values. We incorporate this ambiguity in the value of the mixing coefficient in the form of a systematic uncertainty. Our $\mathcal{O}(a)$-corrected value for $d_2^{(p)}$ decreases by $9$\% when using the LPT estimates for the mixing coefficient when compared to the NPR calculated value. As the aim of this work is to determine the $t$-behaviour of the $\Phi_i$ form factors, rather than a precision determination of $d_2^{(p)}$, we will take the NPR value of the mixing coefficient and quote a further $5$\% systematic uncertainty on our $\mathcal{O}(a)$-improved value of $d_2^{(p)}$.
\newline

The multiplicative renormalisation constant $Z^{[5]}$ is calculated using the procedure outlined in Refs. \cite{Martinelli1994, Gockeler1998}. The scale $\mu^2$ for the renormalisation is set by the squared four-momentum of the quark propagator. Then following the procedure defined in Ref. \cite{Burger2022}, we define a reference scale of $\mu_0 = 2$ GeV and compute the ratio $Z^{[5]}(\mu^2 = p^2)/Z^{[5]}(\mu^2 = \mu_0^2)$. Figure \ref{fig: Z5 ratio} depicts this ratio for all three ensembles. Below this reference scale, the behaviour of all three ensembles appears identical, whereas above this scale, the behaviour diverges. This ratio should have a continuum limit, and so we extrapolate the value of this ratio from our three lattice spacings to $a = 0$ using a quadratic polynomial in $a^2$. Denoting the $a\to 0$ extrapolated ratio as $R(\mu,\mu_0)$, the renormalisation constant at some scale $\mu'$ is then computed from
\begin{equation}
    Z^{[5]}(\mu',a) = R(\mu',\mu_0) Z^{[5]}(\mu_0,a).
\end{equation}
\newline

\begin{figure}[t]
    \centering
    \includegraphics[width = 8.6cm]{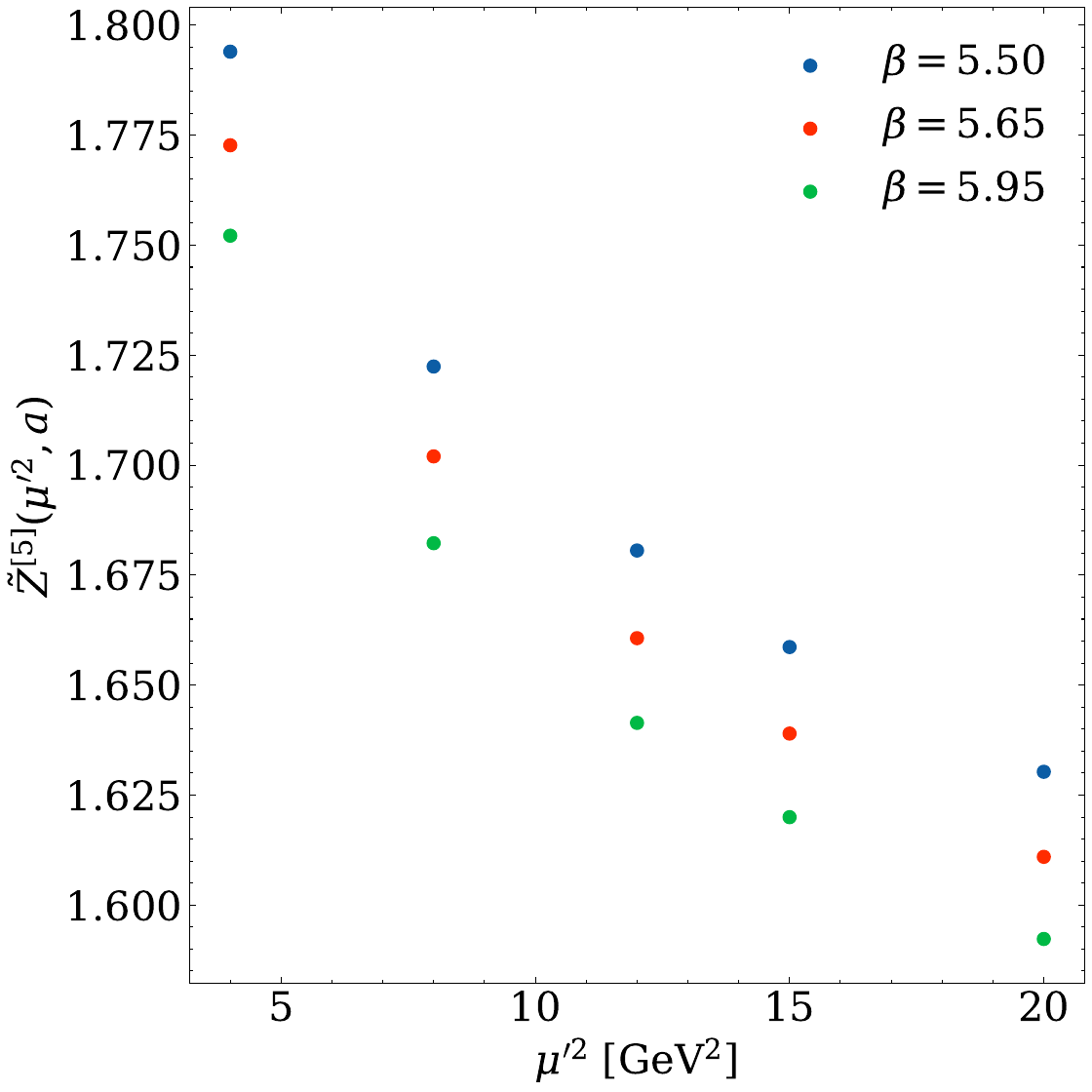}
    \caption{Intermediate scale $\mu'$ dependence of the overall renormalisation factor $\Tilde{Z}(\mu_0,a)$.}
    \label{fig: Z intermediate scale dep}
\end{figure}

The renormalisation in the RI$^\prime$-MOM scheme is performed at an intermediate scale $\mu'=2$ GeV, to match the process followed in Ref. \cite{Burger2022}. The results are then evolved to a common scale of $\mu = 2$ GeV through the one-loop formula for flavour non-singlet operators,
\begin{equation}
    \Phi_i(t, \mu) = \left( \frac{\alpha_s(\mu')}{\alpha_s(\mu)} \right)^{-B} \Phi_i(t, \mu'),
\end{equation}
where $\alpha_s(\mu)$ is the strong coupling constant at the scale $\mu$ and
\begin{equation}
    B = \frac{1}{\frac{11}{3}N_c - \frac{2}{3}N_f}\left(3N_c - \frac{1}{6}\left(N_c - \frac{1}{N_c}\right) \right)
\end{equation}
with $N_c = 3$ and $N_f = 3$. We perform a study of the intermediate scale $\mu'$ dependence to assess any systematic uncertainty introduced through our choice of $\mu'$. The overall renormalisation factor $\Tilde{Z}(\mu^\prime,a)$ at various intermediate scales $\mu^\prime$ is shown in Figure \ref{fig: Z intermediate scale dep}. We find that the choice of intermediate scale introduces a variation of up to 8\% if we were to pick a larger intermediate scale. We incorporate this effect as a systematic uncertainty of 8\% in our estimate of $d_2^{(p)}$ and $d_2^{(n)}$.
\newline

\begin{figure}[t]
    \centering
    \includegraphics[width = 8.6cm]{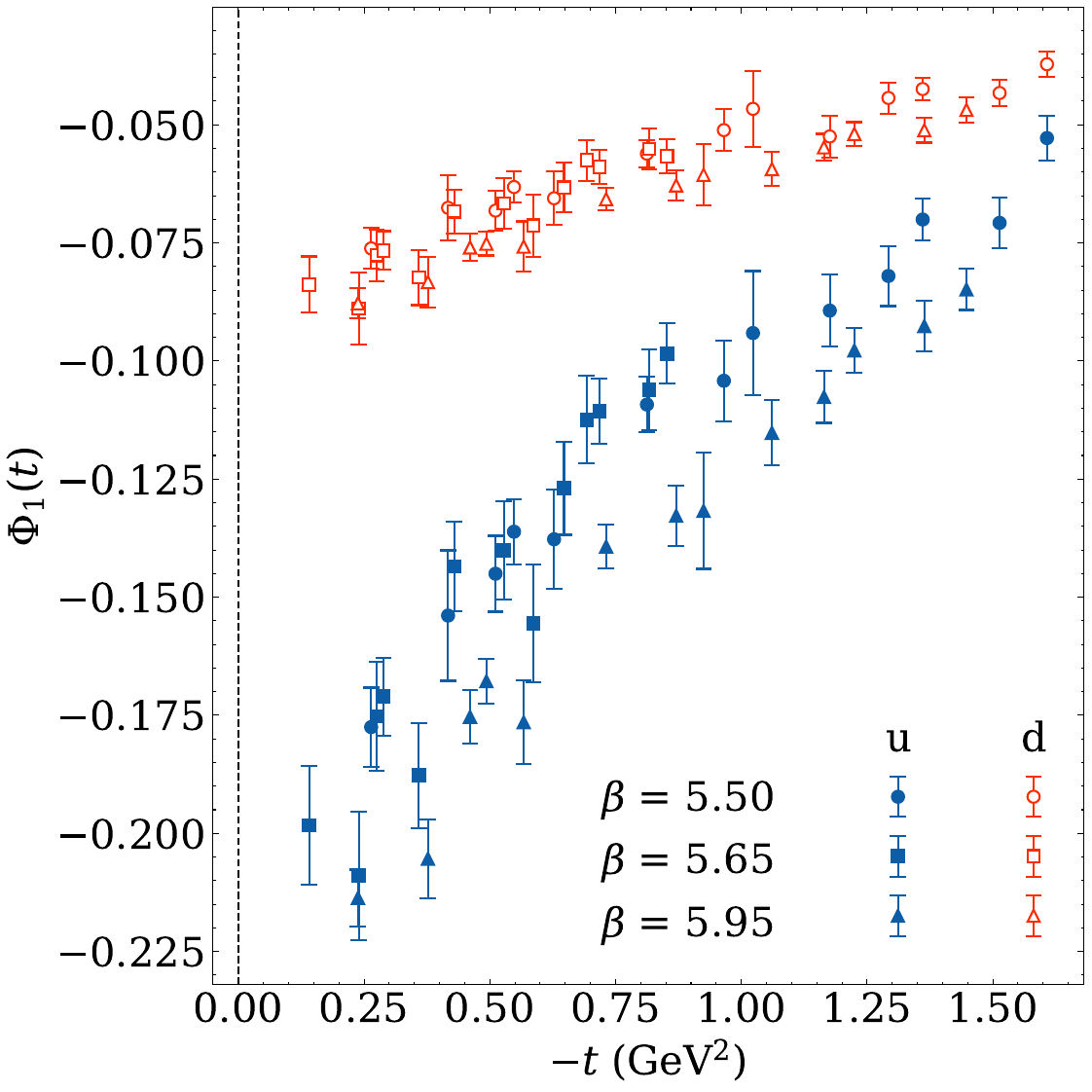}
    \caption{Bare results for the $\Phi_1$ form factor on all lattices for both up and down quarks.}
    \label{fig: bare Phi 1}
\end{figure}

\begin{figure}[t]
    \centering
    \includegraphics[width = 8.6cm]{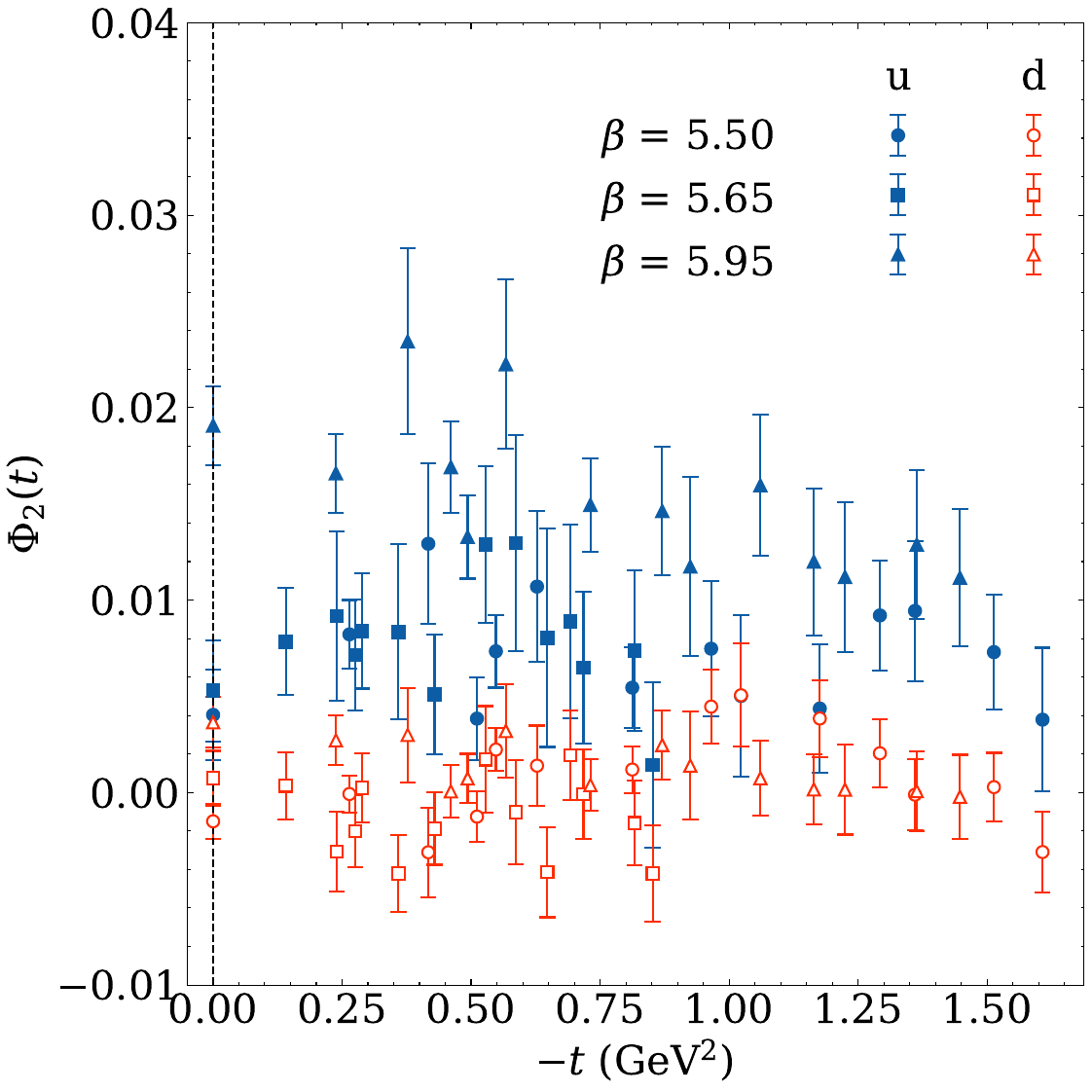}
    \caption{Bare results for the $\Phi_2$ form factor on all lattices for both up and down quarks.}
    \label{fig: bare Phi 2}
\end{figure}

\begin{figure}[t]
    \centering
    \includegraphics[width = 8.6cm]{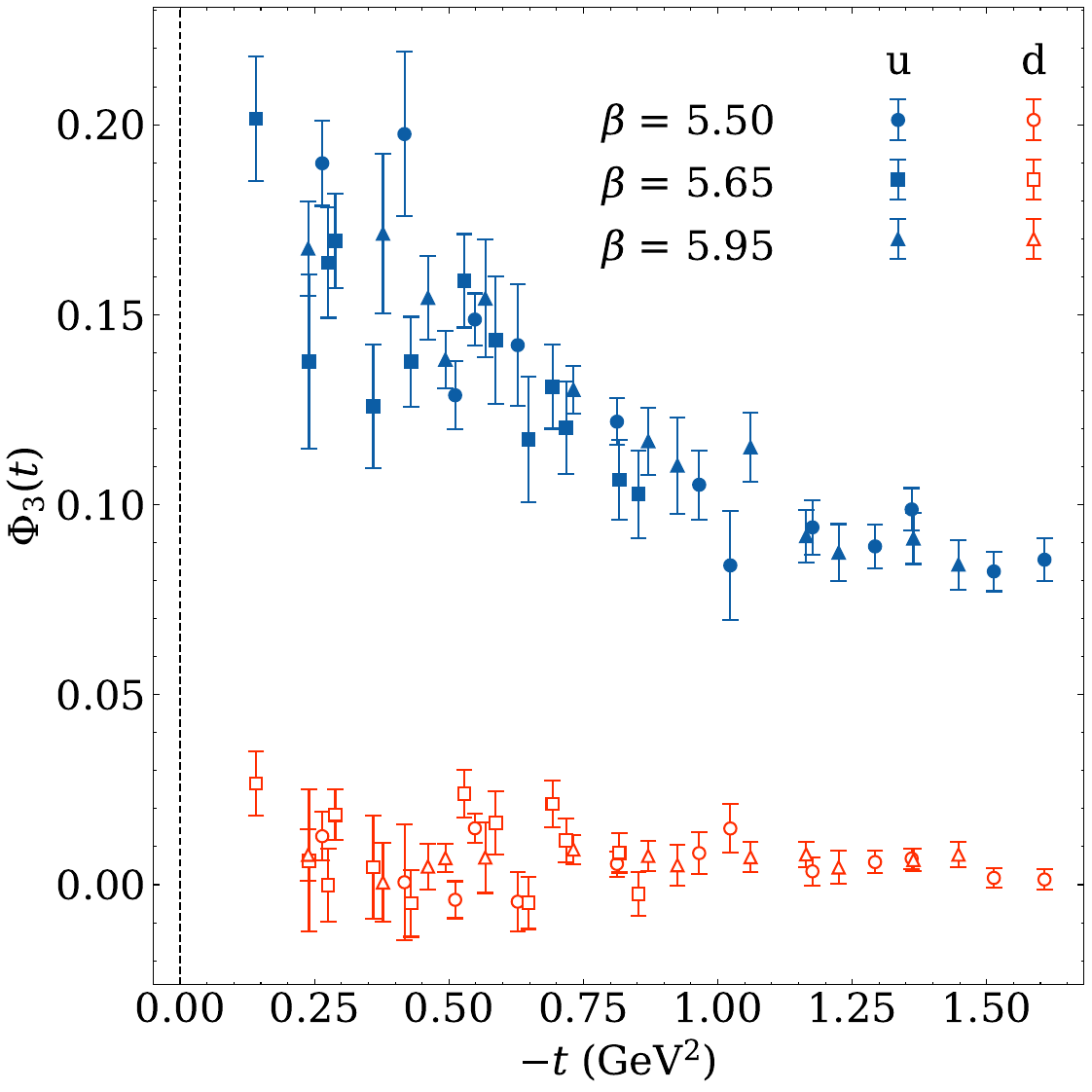}
    \caption{Bare results for the $\Phi_3$ form factor on all lattices for both up and down quarks.}
    \label{fig: bare Phi 3}
\end{figure}

To illustrate the impact of the renormalisation procedure on the form factors, we include the bare form factor results in Figures \ref{fig: bare Phi 1}, \ref{fig: bare Phi 2} and \ref{fig: bare Phi 3}.

\subsection{Discretisation Artefacts}
\begin{figure}[t]
\centering
\begin{minipage}[t]{.49\textwidth}
  \centering
    \includegraphics[width = \textwidth]{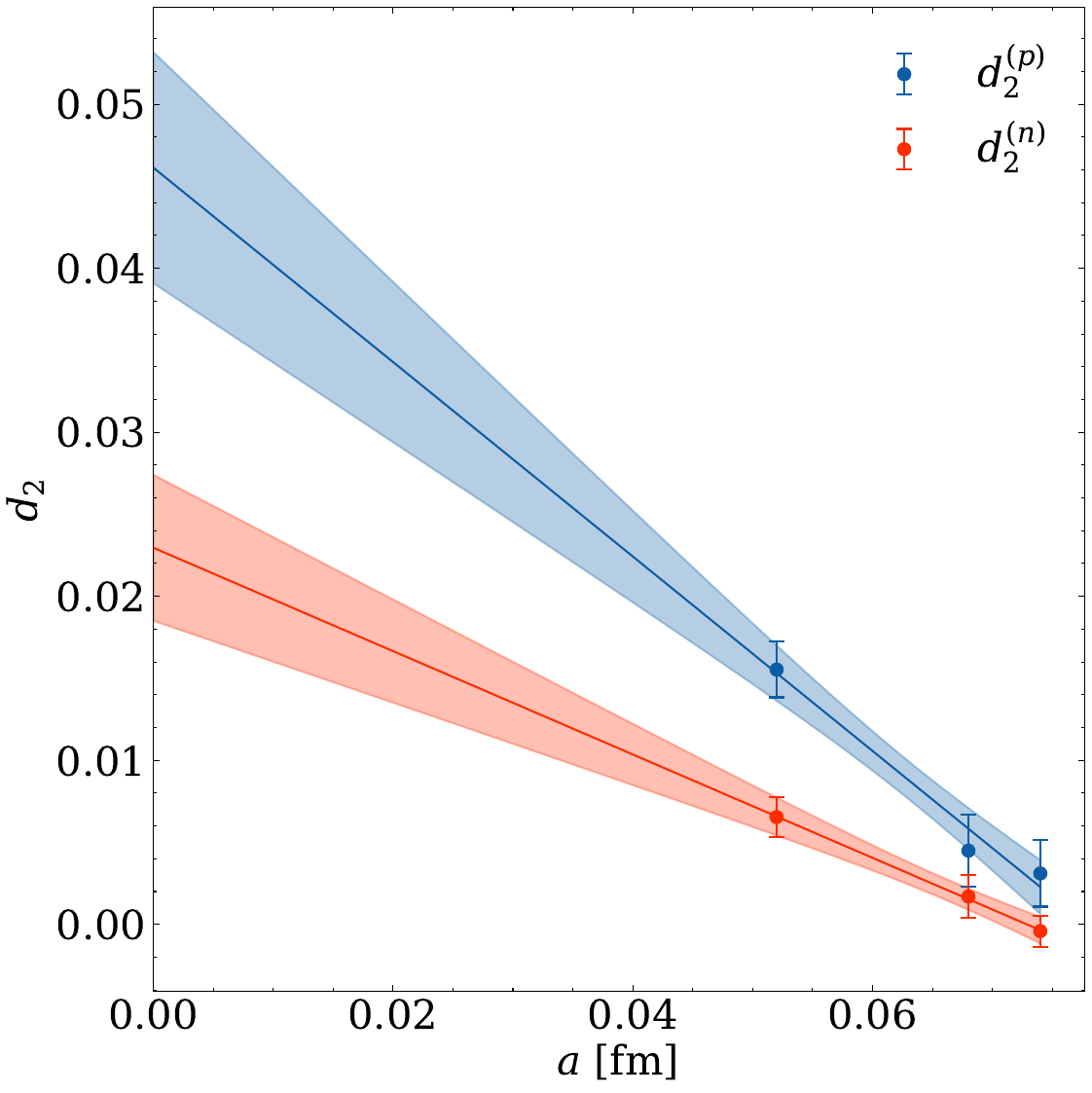}
    \caption{Lattice spacing parameterisation in $a$ for the twist-three forward matrix elements $d_2^{(p)}$ and $d_2^{(n)}$.}
    \label{fig: d2 a extrap}
\end{minipage}%
\hfill
\begin{minipage}[t]{.49\textwidth}
    \centering
    \includegraphics[width = \textwidth]{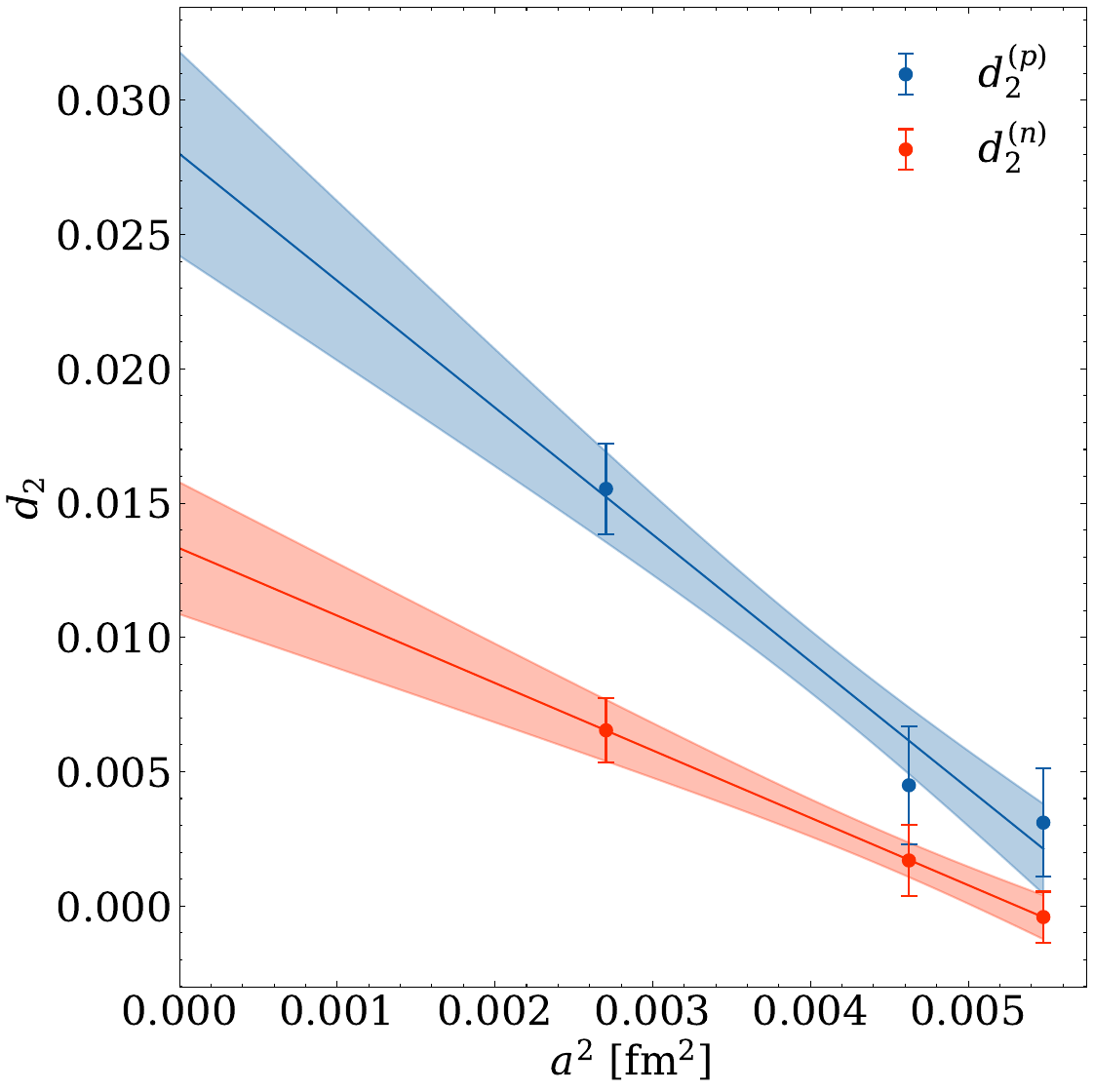}
    \caption{Lattice spacing parameterisation in $a$ for the twist-three forward matrix elements $d_2^{(p)}$ and $d_2^{(n)}$.}
    \label{fig: d2 a2 extrap}
\end{minipage}
\end{figure}

The quark-level results for $d_2$ can be related to the proton and neutron values by
\begin{equation}
    d_2^{(p)} = \left( \frac{2}{3}\right)^2d_2^{(u)} + \left(-\frac{1}{3} \right)^2 d_2^{(d)},
\end{equation}
\begin{equation}
    d_2^{(n)} = \left( \frac{2}{3}\right)^2d_2^{(d)} + \left(-\frac{1}{3} \right)^2 d_2^{(u)}.
\end{equation}
The nucleon quantities $d_2^{(p)}$ and $d_2^{(n)}$ are extrapolated linearly both in $a$ and $a^2$ to $a = 0$. These extrapolations are shown in Figures \ref{fig: d2 a extrap} and \ref{fig: d2 a2 extrap}. Due to the limited number of lattice spacings in this study, both a linear and quadratic extrapolation in the lattice spacing fit the data well, and so we incorporate this as a systematic uncertainty in our $a=0$ estimate for $d_2^{(p/n)}$. The extrapolation in $a$ yields $d_2^{(p)}(m_\pi\approx 450\, \rm{MeV}, a =0 )=0.046(7)_{\rm{stat}}$ and $d_2^{(n)}(m_\pi\approx450\,\rm{MeV}, a=0) = 0.023(5)_{\rm{stat}}$, while the extrapolation in $a^2$ yields $d_2^{(p)}(m_\pi \approx 450\,\rm{MeV},a=0)=0.028(4)_{\rm{stat}}$ and $d_2^{(n)}(m_\pi\approx 450\,\rm{MeV},a=0)=0.0133(3)_{\rm{stat}}$. The extrapolation in $a^2$ represents a decrease of $39$\% in our estimate for $d_2^{(p)}(a=0)$ and a decrease of $42$\% in our estimate for $d_2^{(n)}(a=0)$, and so we add an additional $20$\% systematic uncertainty to both of our estimates. We quote our extrapolation in $a$ in the main text to be consistent with previous calculations\cite{Gockeler2005, Burger2022}. As our ensembles only cover a very limited pion mass range, we do not attempt an extrapolation towards the physical pion mass. Similarly, as all three ensembles use degenerate quark masses at the SU(3) symmetric point, we do not attempt a flavour-symmetry breaking extrapolation. \newline

\begin{figure}[t]
    \centering
    \includegraphics[width = 8.6cm]{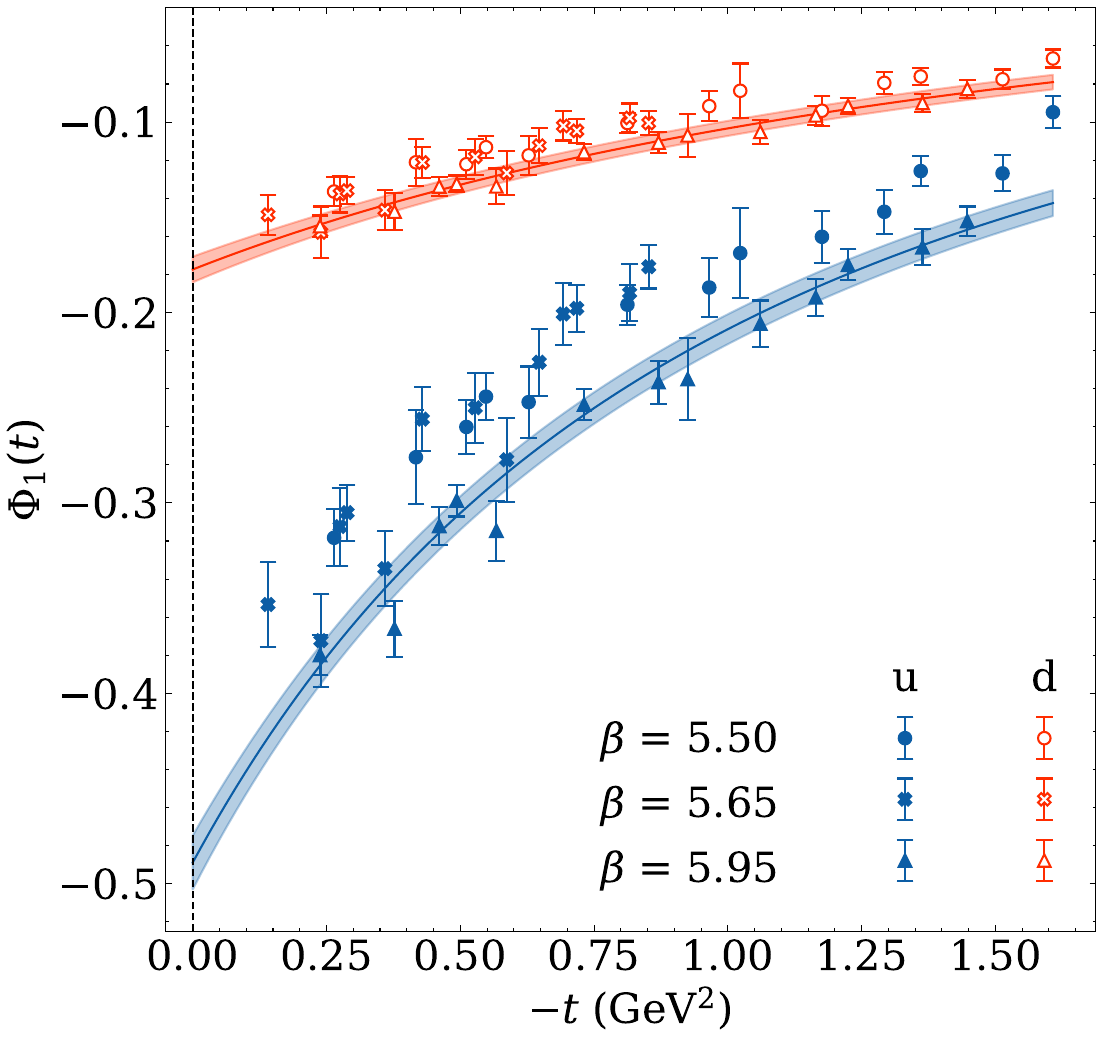}
    \caption{Renormalised lattice results for the $\Phi_1$ form factor for both up and down quarks.}
    \label{fig: Phi1 raw}
\end{figure}

\begin{figure}[t]
    \centering
    \includegraphics[width = 8.6cm]{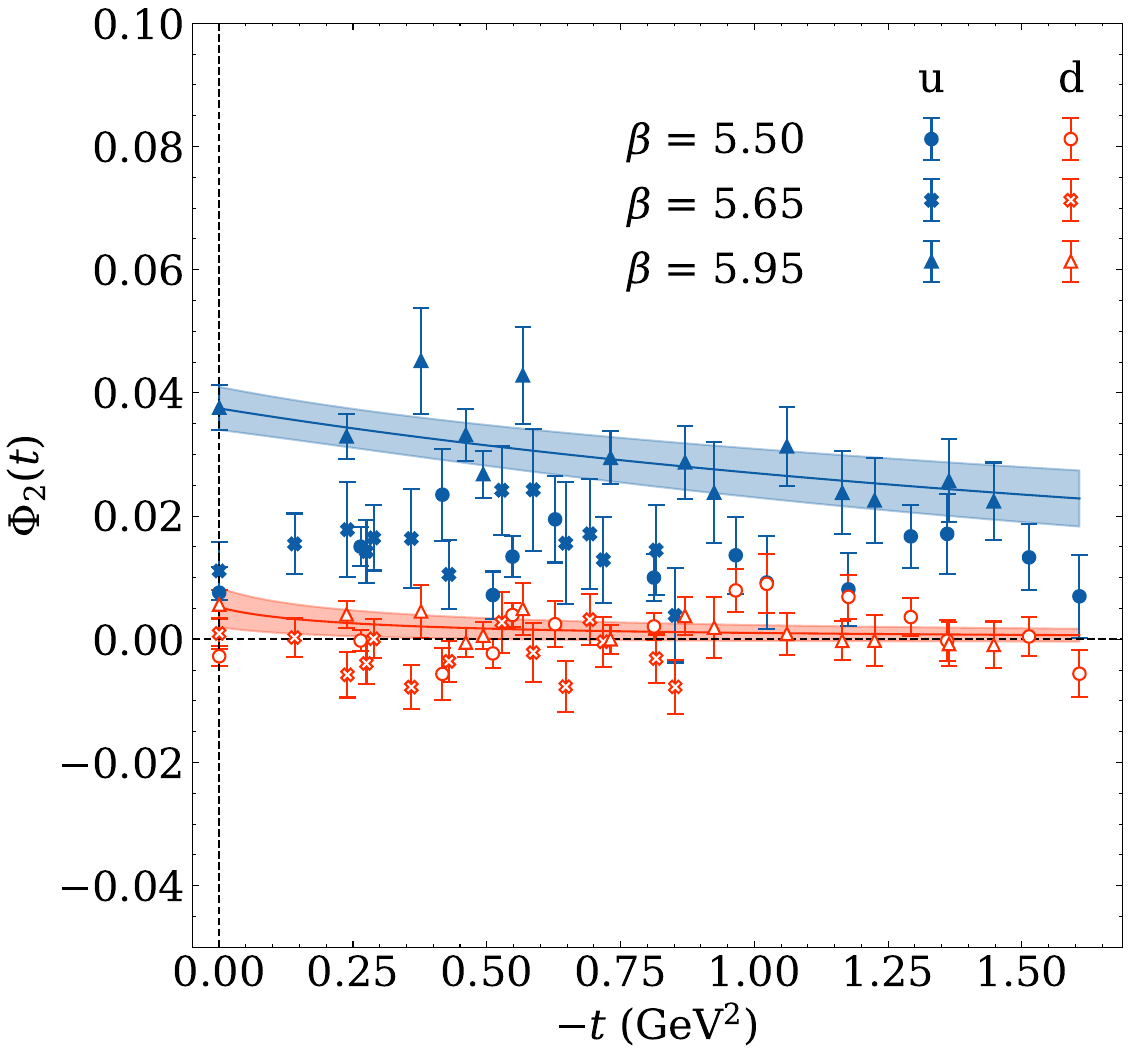}
    \caption{Renormalised lattice results for the $\Phi_2$ form factor for both up and down quarks.}
    \label{fig: Phi2 raw}
\end{figure}

\begin{figure}[t]
    \centering
    \includegraphics[width = 8.6cm]{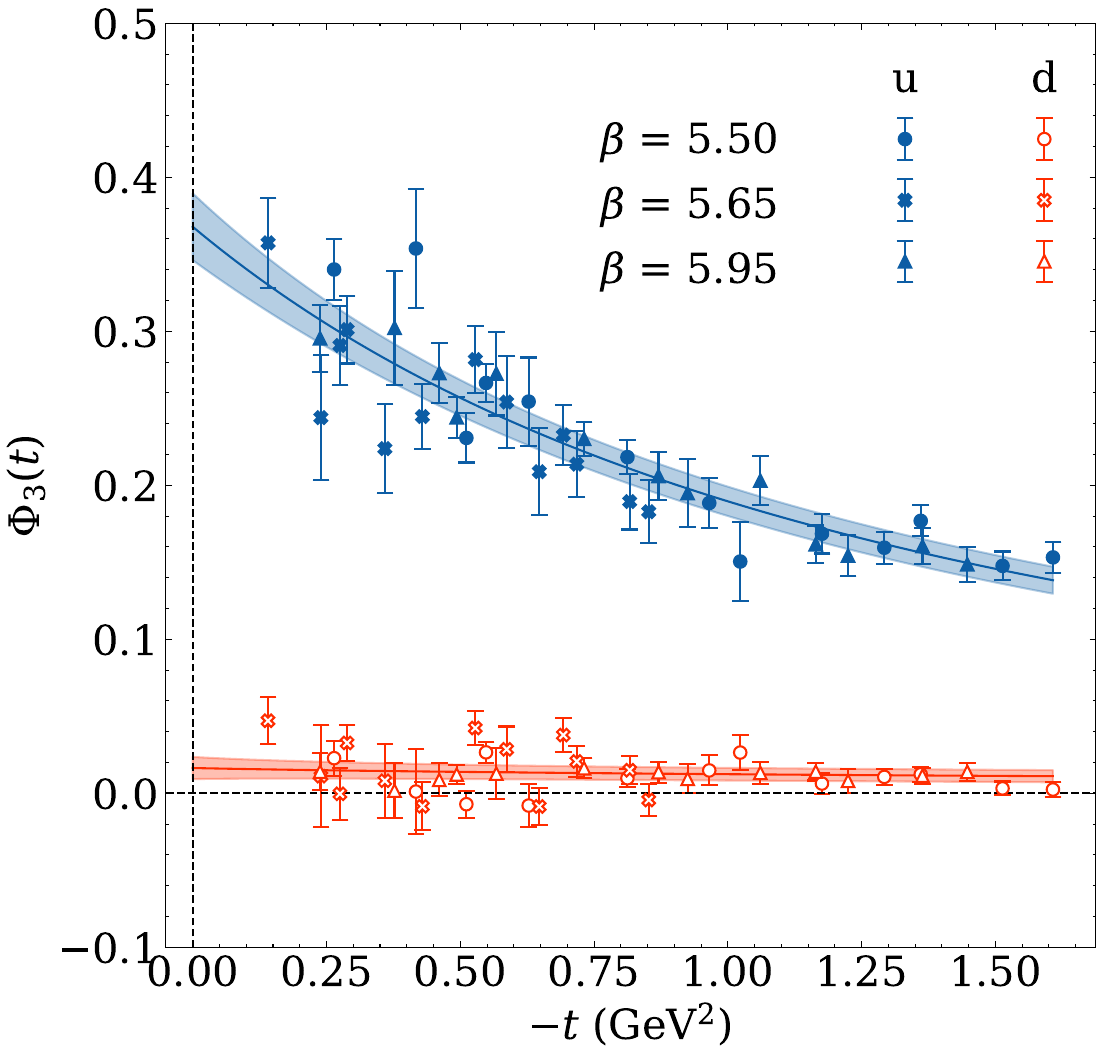}
    \caption{Renormalised lattice results for the $\Phi_3$ form factor for both up and down quarks.}
    \label{fig: Phi3 raw}
\end{figure}

 The renormalised lattice results for the form factors $\Phi_1(t)$, $\Phi_2(t)$ and $\Phi_3(t)$, prior to the subtraction of discretisation artefacts, are shown in Figures \ref{fig: Phi1 raw}, \ref{fig: Phi2 raw} and \ref{fig: Phi3 raw} respectively. The $a$-dependence is most pronounced in Figure \ref{fig: Phi1 raw}, where the finest lattice spacing result ($\beta = 5.95$) is clearly distinct from the other two ensembles for the up quark. Similarly, the $a$-dependence is visible in Figure \ref{fig: Phi2 raw}, however due to the smaller magnitude of $\Phi_2$ relative to $\Phi_1$, this effect is not as pronounced. The lattice spacing appears to have negligible effect on the $\Phi_3$ form factor, as shown in Figure \ref{fig: Phi3 raw}. In all three plots, a dipole function is fit to the $\beta = 5.95$ data to guide the eye.
\newline

We introduce $a$-dependent modifications to the fit parameters of the dipole model,
\begin{equation}
    \Phi_i(a,t) = \frac{\Phi_i(0) + b_i a }{\left(1 - t\left(\frac{1}{\Lambda_i^2} + c_i a  \right) \right)^2},
\end{equation}
where the parameters $b_i,c_i$ are determined through a global fit to the $\Phi_i$ data across all three lattice spacings. We find that corrections to both the dipole mass $\Lambda_i$ and charge $\Phi_i(0)$ are required to capture the continuum behaviour for $\Phi_1(t)$, however only corrections to the charge are required for $\Phi_2(t)$ and $\Phi_3(t)$, i.e. we choose $c_2 = c_3 = 0$. The raw lattice results have their discretisation artefacts removed according by subtracting off the difference from the continuum fit,
\begin{equation}
    \Phi_i^{new} = \Phi_i^{raw} + (\Phi_i^{fit}(0,t) - \Phi_i^{fit}(a,t)).
\end{equation}

\subsection{Impact Parameter Space Representation of Force Densities}
Beginning with Equation (8) from the main text and working in Minkowski space, we focus on each of the contributions one at a time. The first term can then be written as
\begin{equation}
\label{eq: F1 IPS step 1}
    \mathcal{F}_{1,s's}^i(\mathbf{b}) = \frac{i}{\sqrt{2}}\int \frac{d^2\Delta}{(2\pi)^2} e^{-i\mathbf{b}\cdot \mathbf{\Delta}}\overline{u}(p',s')[\Delta^i \gamma^+ \Phi_1(t)]u(p,s).
\end{equation}
The off-forward Dirac bilinear can be computed using the front-form expressions derived in Ref. \cite{Lorce2018}. In order to develop probability interpretations in impact parameter space, we impose $\Delta^+ = 0$. For Equation \eqref{eq: F1 IPS step 1}, it then follows that $\overline{u}(p',s')\gamma^+ u(p,s) = 2P^+\delta_{s's}$. We can then replace the $\Delta^i$ factor with a derivative with respect to $b_i$ acting on the exponential. This yields
\begin{equation}
    \begin{split}    
    \mathcal{F}_{1,s's}^i(\mathbf{b}) &= \frac{2iP^+}{\sqrt{2}}\int\frac{d^2 \Delta}{(2\pi)^2}\frac{1}{-i}\frac{\partial}{\partial b_i}e^{-i\mathbf{b}\cdot \mathbf{\Delta}}\Phi_1(t)\delta_{s's}\\
    &= - 2\sqrt{2}P^+ b^i \frac{\partial}{\partial b^2} \Tilde{\Phi}_1(b^2)\delta_{s's}\\
    &= -2\sqrt{2}P^+ b^i \Tilde{\Phi}_1'(b^2)\delta_{s's},
    \end{split}
\end{equation}
where $\Tilde{\Phi}_1(b^2)$ is the 2D Fourier transform of $\Phi_1(t)$. Next, we consider the second contribution,
\begin{equation}
    \mathcal{F}_{2,s's}^i(\mathbf{b}) = \frac{i}{\sqrt{2}}\int \frac{d^2 \Delta}{(2\pi)^2} e^{-i\mathbf{b}\cdot \mathbf{\Delta}}\overline{u}(p',s')[m_N i\sigma^{+i}]u(p,s).
\end{equation}
Similarly to before, we compute the required off-forward Dirac bilinear in front-form, yielding $\overline{u}(p',s')i\sigma^{+i}u(p,s) = -i\epsilon^{+i\mu\nu}P_\mu (S_\nu)_{s's}$, where $S_{s's}^\mu$ is the average covariant spin-density matrix. Given we are considering a transversely polarised proton travelling in the $\hat{z}$ direction, we take $s_z = 0$. These conditions reduce the contractions in the 4D Levi-Civita symbol down to a 2D subspace,
\begin{equation}
    \begin{split}
        -i\epsilon^{+i\mu\nu}P_\mu S_\nu &= -i\epsilon^{ij}P_-(S_j)_{s's}\\
        &= -\frac{m_N}{\sqrt{2}}i\epsilon^{ij}(S_j)_{s's}.\\
    \end{split}
\end{equation}
The expression for $\mathcal{F}_{2,s's}^i(\mathbf{b})$ then becomes
\begin{equation}
\begin{split}
    \mathcal{F}_{2,s's}^i(\mathbf{b}) &= \frac{i}{\sqrt{2}}\int \frac{d^2 \Delta}{(2\pi)^2}e^{-i\mathbf{b}\cdot\mathbf{\Delta}}m_N \left[-\frac{m_N}{\sqrt{2}}i\epsilon^{ij}(S_j)_{s's}\right]\Phi_2(t)\\
    &= m_N^2\epsilon^{ij}(S_j)_{s's} \Tilde{\Phi}_2(b^2).    
\end{split}
\end{equation}
Finally, we consider the third contribution,
\begin{equation}
    \mathcal{F}_{3,s's}^i(\mathbf{b}) = \frac{i}{\sqrt{2}}\int \frac{d^2\Delta}{(2\pi)^2} e^{-i\mathbf{b}\cdot \Delta} \\
    \overline{u}(p',s')\bigg[\frac{\Delta^i}{m_N}i\sigma^{+\Delta}\Phi_3(t) \bigg]u(p,s).
\end{equation}
Similarly to $\mathcal{F}_{2,s's}^i(\mathbf{b})$, the Dirac bilinear is 
\begin{equation}
    \overline{u}(p',s')i\sigma^{+\Delta}u(p,s) = -i\epsilon^{+\Delta \mu\nu}P_\mu (S_\nu)_{s's},
\end{equation}
which can also be reduced to a 2D subspace,
\begin{equation}
    -i\epsilon^{+\Delta \mu\nu}P_\mu (S_{\nu})_{s's} = -i\epsilon^{jk}\Delta_j (S_k)_{s's}.
\end{equation}
The expression for $\mathcal{F}_{3,s's}^i(\mathbf{b})$ is then
\begin{equation}
    \mathcal{F}_{3,s's}^i(\mathbf{b}) = -\epsilon^{jk}(S_k)_{s's}\int \frac{d^2\Delta}{(2\pi)^2} e^{-i\mathbf{b}\cdot \Delta} \Delta^i \Delta^j \Phi_3(t),
\end{equation}
where we can then replace the $\Delta^i$ and $\Delta^j$ with derivatives as we did for $\mathcal{F}_{1,s's}^i(\mathbf{b})$, yielding
\begin{equation}
    \mathcal{F}_{3,s's}^i(\mathbf{b}) = -\epsilon^{jk}(S_k)_{s's}\frac{\partial}{\partial b_i}\frac{\partial}{\partial b_j}\Tilde{\Phi}_3(b^2).
\end{equation}
The action of the two derivatives can be recast in terms of derivatives with respect to $b^2$ by careful application of the product rule,
\begin{equation}
    \frac{\partial}{\partial b_i}\frac{\partial}{\partial b_J}\Tilde{\Phi}_3(b^2) = 2\delta_{ij}\Tilde{\Phi}'_3(b^2) + 4b_i b_j \Tilde{\Phi}''_3(b^2),
\end{equation}
leaving us with the final expression for $\mathcal{F}_{3,s's}^i(\mathbf{b})$,
\begin{equation}
    \mathcal{F}_{3,s's}^i(\mathbf{b}) = -\epsilon^{jk}(S_k)_{s's}[2\delta_{ij}\Tilde{\Phi}'_3(b^2) + 4b_i b_j \Tilde{\Phi}''_3(b^2)].
\end{equation}
Therefore, the total force density in impact parameter space can be written as
\begin{equation}
    \mathcal{F}_{s's}^i(\mathbf{b}) = -2\sqrt{2}P^+b^i \Tilde{\Phi}_1'(b^2)\delta_{s's} + m_N^2\epsilon^{ij}\Tilde{\Phi}_2(b^2)(S_j)_{s's} - \epsilon^{jk}[2\delta_{ij}\Tilde{\Phi}_3'(b^2)+4b_i b_j \Tilde{\Phi}_3''(b^2)](S_k)_{s's}.
\end{equation}

\subsection{Model Dependence of Force Magnitude Estimates}
In order to assess the magnitudes of these colour-Lorentz forces, we divide out the quark density dependence along the radial direction. We calculate the quark density from the electromagnetic form factors $F_1(t)$ and $F_2(t)$ using the method outlined in Refs. \cite{QCDSF2006, Diehl2005}. For the force related to the $\Phi_1$ form factor, as there is an axial symmetry and this force profile is isotropic. However, due to the asymmetric force distribution arising from the $\Phi_3$ form factor, as well as asymmetric quark distributions in a polarised nucleon, we choose to study the force profile along the $b_x$ axis for simplicity. The form factor fits shown in the main text make use of a dipole fit function, however for these model dependence studies, we generalise this fit to an $n$-order pole model,
\begin{equation}
    \Phi_i(t) = \frac{\Phi_i(0)}{\left(1 + \frac{t}{\Lambda_i^2} \right)^n},
\end{equation}
with $n=2,3$ and $4$. These fit models approximate the data with comparable $\chi^2/$d.o.f to the dipole model for all $\Phi_i$ form factors. For the corresponding quark densities, we keep the fit model consistent as a dipole.
\newline

\begin{figure}
\centering
    \centering
    \includegraphics[width = 8.6cm]{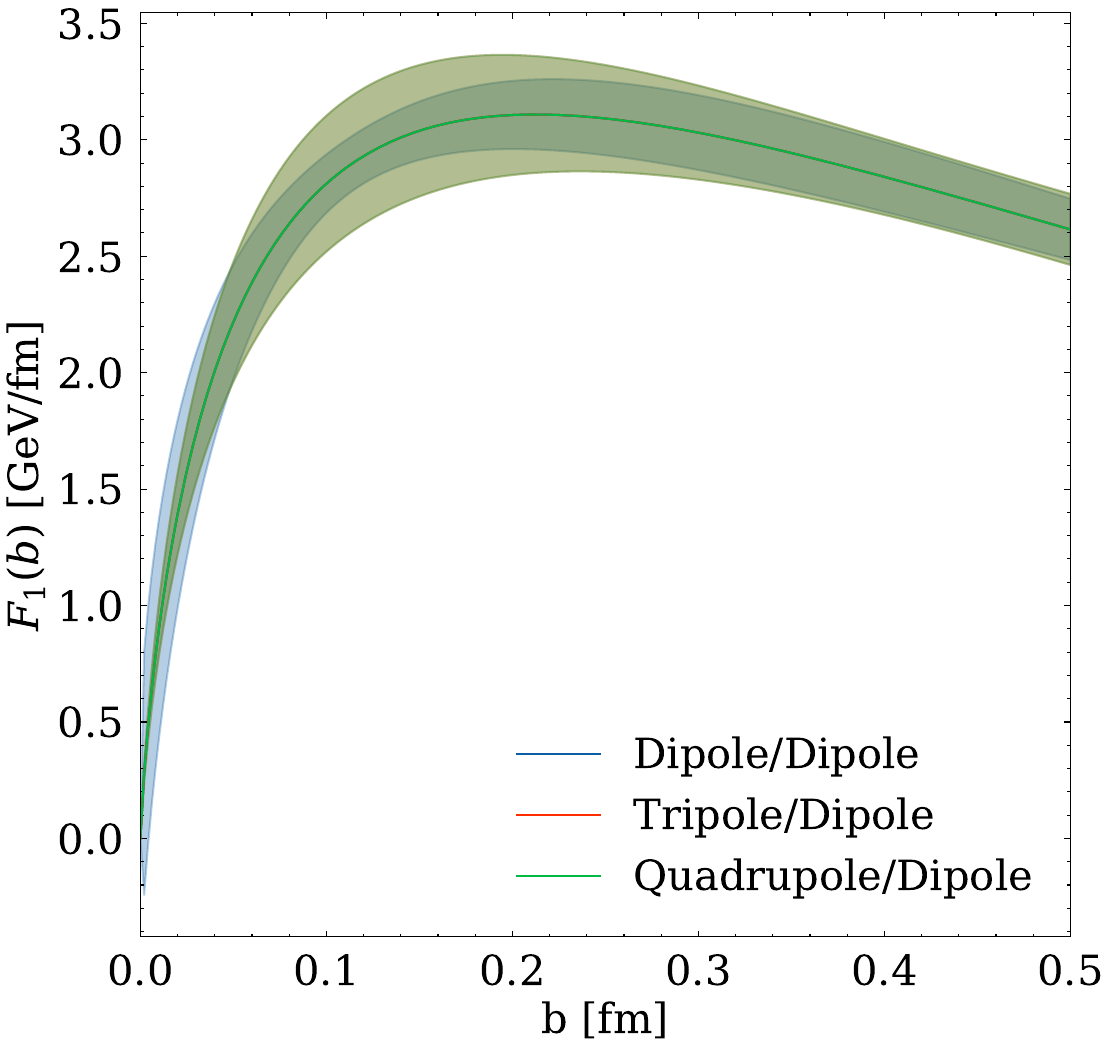}
    \caption{Model dependence of the force magnitude estimate for the colour-Lorentz force corresponding to the $\Phi_1$ form factor.}
    \label{fig: phi1 model dep}
\end{figure}

In Figure \ref{fig: phi1 model dep}, we show the resulting force profile of the colour-Lorentz force corresponding to the $\Phi_1$ form factor as we vary the fit model. Both the quark density and colour-Lorentz force density are computed at our finest lattice spacing. The force magnitude appears insensitive to the model choice for all $b$. The force vanishes at the origin, before rising to a maximum of approximately 3 GeV/fm and remaining relatively constant at larger $b$. Therefore the struck up quark experiences a large attractive force towards the nucleon centre of mass the instant it is struck by the virtual photon. Given the limited $Q^2$ range probed in this study, the agreement between the models at small $b$ is quite remarkable.
\newline

\begin{figure}
    \centering
    \includegraphics[width = 8.6cm]{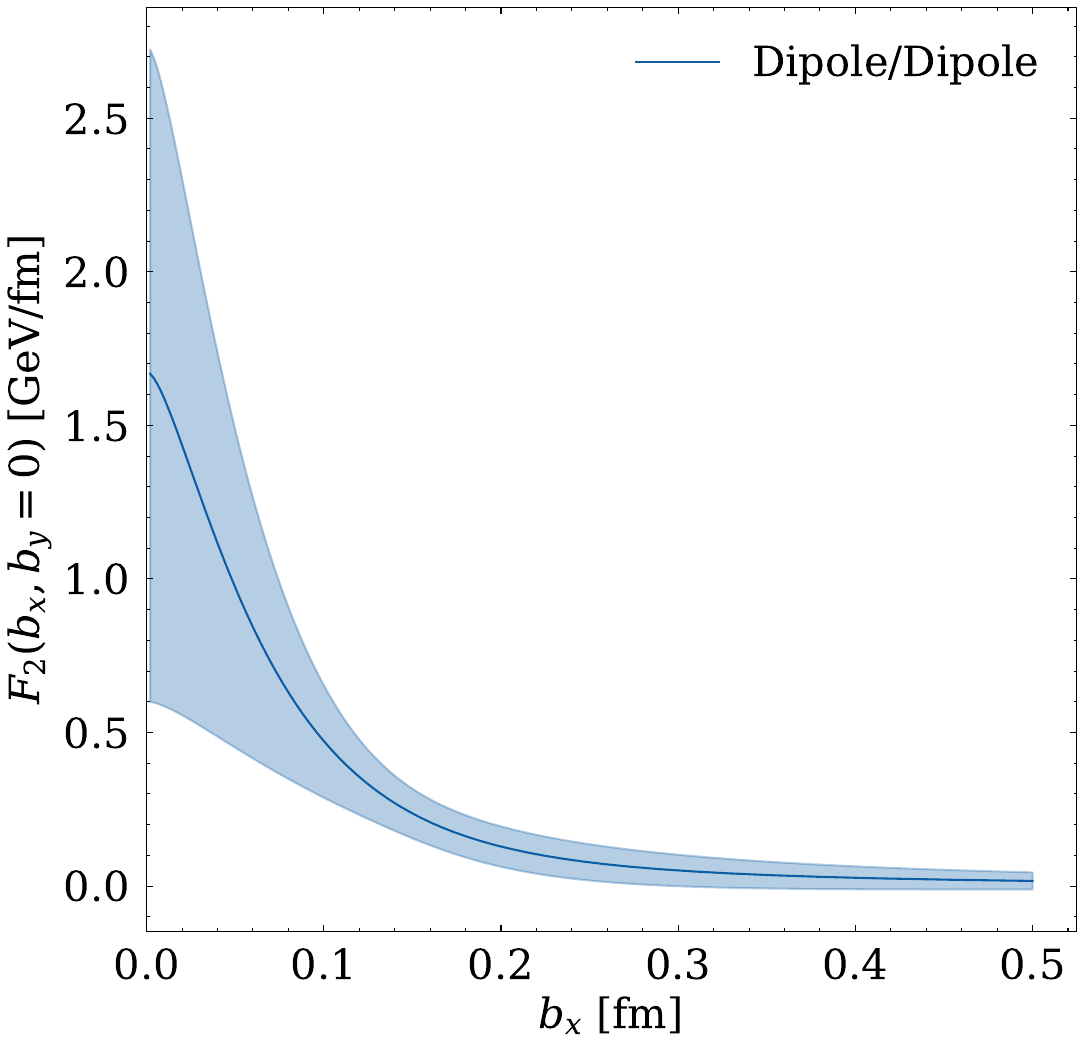}
    \caption{Force magnitude estimate for the colour-Lorentz force corresponding to the $\Phi_2$ form factor.}
    \label{fig: phi2 model dep}
\end{figure}

The force profile of the colour-Lorentz force corresponding to the $\Phi_2$ form factor is shown in Figure \ref{fig: phi2 model dep}. Due to large fluctuations in the fit parameters for both the tripole and quadrupole fits to the form factor, these have been omitted. Better control of the statistical fluctutations of $\Phi_2$ would be required before these fits are attempted. Furthermore, the larger uncertainty at small $b$ is a consequence of uncertain behaviour at larger momentum transfers. Due to the smaller magnitude of the $\Phi_2$ form factor, the force resulting from the Fourier transform of this form factor is also smaller in magnitude. The force profile also decays at a faster rate when compared to the other colour-Lorentz forces.\newline

\begin{figure}
    \centering
    \includegraphics[width = 8.6cm]{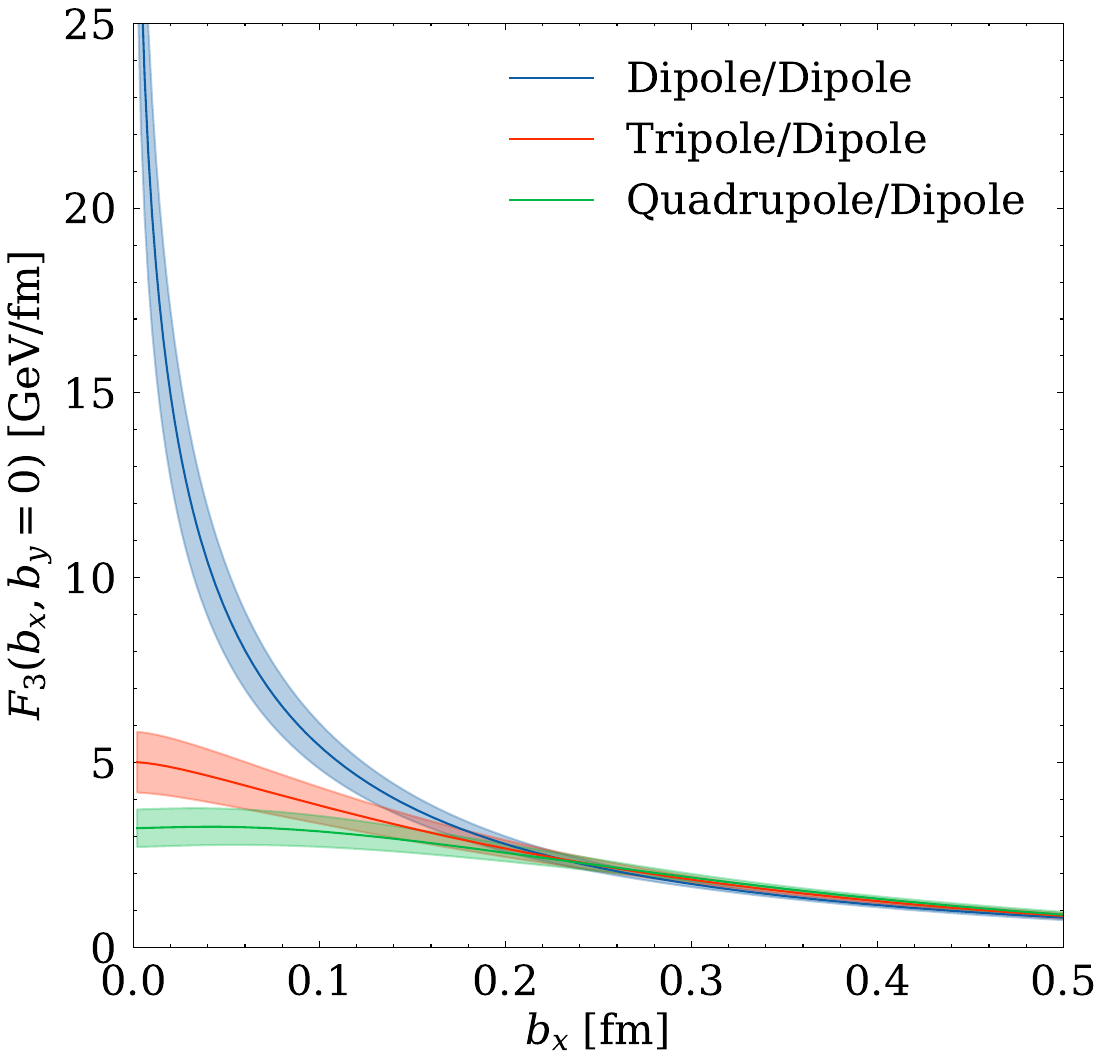}
    \caption{Model dependence of the force magnitude estimate for the colour-Lorentz force corresponding to the $\Phi_3$ form factor.}
    \label{fig: phi3 model dep}
\end{figure}

Figure \ref{fig: phi3 model dep} shows the resulting force profile of the colour-Lorentz force corresponding to the $\Phi_3$ form factor as we vary the fit model. Both the quark density and colour-Lorentz force density are computed at our finest lattice spacing. Compared to the isotropic restoring force due to $\Phi_1$, the spin-dependent force shows significantly more model dependence at small $b$. The dipole fit to the $\Phi_3$ form factor results in a singularity at the origin. The higher-order pole fits, whilst remaining finite at the origin, remain large close to the origin. The force magnitude becomes model independent at a distance of approximately 0.25 fm, likely owing to the low $Q^2$ region probed by our study. The low $b$ region could be further constrained by probing higher momentum transfers, however this would likely result in considerably noisier signals for the form factors. 

